\documentclass[11pt,a4paper]{article}
\pdfoutput=1
\usepackage{jheppub}
\usepackage{mathrsfs,graphicx,rotating,amsmath,amsfonts,mathtools,booktabs,wasysym}
\usepackage{slashed}
\usepackage[table,xcdraw,dvipsnames]{xcolor}
\usepackage{graphicx}
\usepackage{bbold}
\usepackage[utf8x]{inputenc}
\usepackage[english]{babel}
\usepackage{multirow,multicol}
\usepackage{epstopdf}
\usepackage{bbm}
\usepackage{changepage}
\usepackage{appendix}
\usepackage{systeme}
\usepackage{libertine}
\usepackage{braket}
\usepackage{wasysym}
\usepackage{empheq}
\usepackage{cancel}
\usepackage{enumitem}
\usepackage{mathrsfs}
\usepackage{lmodern}
\usepackage{tabularx}
\usepackage{multicol}
\usepackage{color}
\usepackage{mathtools}
\usepackage{verbatim}
\usepackage{amssymb}
\usepackage{amsfonts}

\newcommand{\be}{\begin{equation}}
\newcommand{\ee}{\end{equation}}
\newcommand{\bea}{\begin{eqnarray}}
\newcommand{\eea}{\end{eqnarray}}

\newcommand{\beqa}{\begin{eqnarray}}
\newcommand{\eeqa}{\end{eqnarray}}

\newcommand{\bigzero}{\mbox{\normalfont\Large\bfseries 0}}
\newcommand{\rvline}{\hspace*{-\arraycolsep}\vline\hspace*{-\arraycolsep}}

\setcitestyle{square}

\DeclarePairedDelimiter\abs{\lvert}{\rvert}%
\title{Attractive (s)axions: cosmological trackers at the boundary of moduli space}
\author[a]{Filippo Revello}
\affiliation[a]{Institute for Theoretical Physics\\ Utrecht University,
Princetonplein 5, 3584 CC Utrecht, The Netherlands}
\emailAdd{f.revello@uu.nl}
\abstract{We study the cosmological evolution of a FLRW universe dominated by the energy density of moduli close to asymptotic regions of moduli space. Due to the structure of the $\mathcal{N}=1$ SUGRA kinetic term, a saxion and an axion residing in the same chiral multiplet are (universally) coupled even if the latter is a flat direction of the potential, resulting in non-trivial dynamics. We generalise known results in the literature to the case of multiple moduli, showing the existence of various ``tracker" attractor solutions where the relative energy densities of many components (axions included) stay in a fixed ratio throughout the evolution. We conclude with some phenomenological applications, relevant for both the early and late universe.}
\begin{document}

\maketitle
\section{Introduction}

A fundamental obstruction to making contact between String Theory and observations is the existence of an incredibly large number of vacua, commonly referred to as the \emph{Landscape} \cite{Susskind:2003kw}, which can vastly differ in their low-energy predictions. It is extremely hard to identify vacua which satisfy all the required properties (stabilised moduli, scale separation, positive cosmological constant, etc) and reproduce the Standard Model (SM) of particle physics; let alone use them to investigate \emph{new} phenomena predicted by String Theory and accessible at low energies. In recent years, this problem has been exacerbated by the realisation that many such vacua may not even, after all, correctly descend from String Theory, but rather belong to a \emph{Swampland} \cite{Vafa:2005ui,Ooguri:2006in,Palti:2019pca} of theories which do not admit a consistent UV completion with gravity. Indeed, the very existence of de Sitter vacua has been cast into doubt \cite{Danielsson:2018ztv,Obied:2018sgi}, and it is not clear whether a positive resolution of this conundrum will require knowledge beyond our current understanding of the theory, for instance in the non-perturbative regime \cite{Dine:1985he,VanRiet:2023pnx}. 

While this outlook might seem fairly bleak, it is tempting to ask whether even with the String Theory that \emph{we do know}, and irrespectively of the precise details of the `true vacuum' we supposedly live in, it is possible to make \emph{some} predictions relevant to phenomenology. This is the fundamental motivation behind this work, and the point of view we wish to explore throughout the rest of the paper. Given the difficulties stated above, it is natural to turn to features of string compactifications which are, more or less, model independent, and to a certain extent holding across different constructions. Perhaps the most general (and striking) prediction of String Theory is the existence of extra dimensions, and all of the associated structures that come with them. Indeed, within a 4d Effective Field Theory (EFT) approach, the geometry of extra dimensions is parametrised by scalar fields, the moduli. Typical compactifications may contain $\mathcal{O}(100)$ of them, and their existence is completely generic. To avoid constraints arising from fifth force experiments, moduli need to be stabilised by the introduction of potentials, which in turn can lead to non-trivial cosmological dynamics, such as a phase of inflation. Moduli stabilisation, together with their cosmological consequences, have been studied intensely (See \cite{Grana:2005jc,Douglas:2006es,VanRiet:2023pnx} and \cite{Cicoli:2023opf} for reviews respectively). While many of the details may often be model dependent, there are some general lessons to be learnt by this interplay of String Theory and Cosmology. As an example, it is a generic fact that moduli are very weakly (gravitationally) coupled to SM fields, and are thus the last particles to decay after inflation, making them a very good candidate for reheating. In turn, this puts severe constraints on their properties, as consistency with BBN requires a lower bound of order $\mathcal{O} \left( \rm{TeV} \right)$ on their mass \cite{Coughlan:1983ci,Banks:1993en,deCarlos:1993wie}. Another (closely related) example of such a `generic' prediction in string theory is the existence of a plethora of axions or axion like particles (ALPs) \cite{Svrcek:2006yi,Conlon:2006tq,Arvanitaki:2009fg,Jaeckel:2010ni,Cicoli:2012sz}, which arise as zero modes of higher form fields. At the perturbative level, they are a flat direction of the potential, so they can be arbitrarily light if the non-perturbative corrections are small enough. In some cases \cite{Svrcek:2006yi,Conlon:2006tq} it is possible to identify one (linear combination) of them with the QCD axion \cite{Peccei:1977hh,Wilczek:1977pj} and solve the strong CP problem, and they can also be fuzzy dark matter candidates \cite{Preskill:1982cy,Abbott:1982af,Dine:1982ah,Hui:2016ltb,Cicoli:2021gss} or act as the inflaton \cite{Silverstein:2008sg,McAllister:2008hb}.
Although the existence of such an `axiverse' can be argued for on general grounds, the requirement that some of the axions be light enough for phenomenology is non-trivial, and can be heavily influenced by moduli stabilisation \cite{Svrcek:2006yi,Conlon:2006tq}. For this reason, we will mostly focus on scenarios where the stabilisation of moduli is best developed, such as KKLT/LVS \cite{Kachru:2003aw,Balasubramanian:2005zx,Conlon:2005ki} in type IIB, or more general situations where their potentials are well understood, such as close to the boundaries of moduli space.

In this paper, we will be concerned with the study of some universal features of the cosmological evolution of moduli and axions, motivated from a variety of string theory examples. Much work, recent \cite{Cicoli:2018kdo,Hebecker:2019csg,ValeixoBento:2020ujr,Cicoli:2020cfj,Cicoli:2020noz,Brinkmann:2022oxy,Cicoli:2021fsd,Cicoli:2021skd,Christodoulidis:2019jsx,Christodoulidis:2021vye,Christodoulidis:2022vww,Rudelius:2022gbz,Calderon-Infante:2022nxb,Cremonini:2023suw,Freigang:2023ogu,Andriot:2023wvg} and less recent \cite{Antoniadis:1988vi,Dvali:1998pa,Choi:1999xn,Kachru:2003sx,Conlon:2005jm,Silverstein:2008sg,McAllister:2008hb,Cicoli:2008gp} (See \cite{Baumann:2014nda,Cicoli:2023opf} for reviews), has been devoted to embedding some desired epochs of cosmology - such as inflation, reheating or quintessence - within a string-theoretic framework. This is by all means a necessary and urgent goal of String Phenomenology, given the theoretical and experimental motivations. Indeed, we will assume throughout the work that inflation did happen, and that there should be an appropriate stringy description of it. However, we shall take a complementary perspective, and ask instead what possible cosmological evolutions would most naturally follow from the basic structure of string compactifications. In particular, we will analyse the coupled dynamics of axions and saxions within the same multiplet in the presence of typical potentials arising within flux compactifications. While the cosmological evolution of saxions with exponential potentials have been studied in detail \cite{Wetterich:1987fm,Copeland:1997et,Ferreira:1997hj,Barreiro:1999zs,Collinucci:2004iw,Calderon-Infante:2022nxb,Shiu:2023nph,Shiu:2023fhb}, less attention has been paid to the corresponding axions, especially for early universe cosmology (See however \cite{Sonner:2006yn,Russo:2018akp,Cicoli:2020cfj,Cicoli:2020noz,Brinkmann:2022oxy,Russo:2022pgo,Christodoulidis:2021vye}). Within an effective $\mathcal{N}=1$ description, axions are paired to the moduli in a chiral multiplet, and their dynamics are necessarily intertwined. A main point this paper will indeed be to explore how the universal structure of the K\"ahler potential close to boundaries of moduli space is able to couple the cosmological evolution of saxions and axions. Our approach will closely parallel the seminal works \cite{Wetterich:1987fm,Copeland:1997et,Ferreira:1997hj}, by reducing the cosmological equations of motion into a first order autonomous system which can be characterised analytically. A more complete characterization of late time scalar field cosmologies involving saxions (but not axions), was also carried out in the recent works \cite{Shiu:2023nph,Shiu:2023fhb}. 

For any cosmological applications, a necessary assumption is that, at some point of the universe's history, the energy density stored in the moduli was (co)dominating over other sources, and thus driving the cosmic evolution. One place where this could be relevant is the late universe: we will indeed examine in detail under which conditions the above cosmologies can exhibit accelerated expansion, as a stepping stone towards quintessence. 
In the early universe, such an assumption can instead be motivated most naturally in the period immediately after inflation, since any energy density present beforehand (such as in radiation or matter) would have been diluted by the rapid expansion. Along the same lines, it was proposed in \cite{Conlon:2022pnx,Apers:2022cyl} that the typically steep moduli potentials would lead to an epoch of early kination after the end of inflation. In this work, we will generalise these same arguments to the presence of axions, and explore consequences both for the dynamics as a whole and for the axions themselves. In particular, we will show how saxion kination will necessarily induce some dynamics for the axions, unless other sources can overtake the energy density before this can happen. 

The paper is organised as follows: Section \ref{sec:mot} reviews the string theoretic scenarios that motivate the work, while the equations governing the cosmological evolution of the (s)axions are analysed in Section \ref{sec:3}. Applications are presented in Section \ref{sec:app}, before we conclude in Section \ref{sec:con}.

\section{String Theory: motivating scenarios}\label{sec:mot}

\subsection{The saxion-axion EFT}

A central motivation for this work is to investigate the cosmological consequences of moduli potentials that can appear in a realistic scenario deriving from string theory, with particular attention to the dynamics of the axions. At low energies, they will be described by an $\mathcal{N}=1$ theory where the saxions and the axions are paired together into chiral multiplets. At the two-derivative level, the bosonic action of an $\mathcal{N}=1$ supergravity theory with $N$ chiral multiples $\Phi^I$ is given by
\begin{equation}
S= \frac{M_{P,d}^2}{2 }\int d^d x \, \sqrt{- g}  \Big\{ \mathcal{R} +\frac{1}{2} G_{I \bar{J}} \, \partial_{\mu} \Phi^I \partial^{\mu} \bar{\Phi}^{\bar{J}} - V(\Phi, \bar{\Phi} )\Big\}.
\end{equation}
Within the usual supergravity description, the target space metric can be derived from the K\"ahler potential as
\begin{equation}
G_{I \bar{J}} = \partial_I \partial_{\bar{J}} K,
\end{equation}
and the potential is given by the usual $\mathcal{N}=1$ formula
\begin{equation}
V(\Phi, \bar{\Phi} )=e^K\left(\sum_{\Phi_I} K^{I \bar{J}} D_I W D_{\bar{J}} \bar{W}-3|W|^2\right).
\end{equation}
The chiral fields $\Phi^I$ can be decomposed into axionic and saxionic components as
\begin{equation}
\Phi^I = s^I+ i \, a^I.
\end{equation}
Denoting them with the same symbol  $\phi^i = \{s^i,a^i\}$, the scalar field equations of motion are
\begin{equation}\label{eq:eom}
\ddot{\phi}^i+ \Gamma^{i}_{j\,k} \dot{\phi}^j \dot{\phi}^k+(d-1)H\dot{\phi}^i+\partial^i V =0,
\end{equation}
where all the contractions and the Christoffel symbols are defined with respect to the real metric $G_{a b}$, to be complemented by the first Friedmann equation
\begin{equation}\label{eq:f1}
\frac{(d-1)(d-2)}{2}H^2 =\frac{1}{2}G_{i j}\partial_{\mu} \phi^i \partial^{\mu} \phi^j-V(\phi^i).
\end{equation}
The second Friedmann equation is not independent from the ones written above, and can be recovered by taking a time derivative of \eqref{eq:f1} and substituting Eqs. \eqref{eq:eom}-\eqref{eq:f1} back into the result.

While general expressions for the metric $G_{I \bar{J}} (\Phi, \bar{\Phi} )$ and the potential $V(\Phi, \bar{\Phi} )$ can be very difficult to calculate, towards the boundaries of moduli space they typically simplify significantly. Our results will apply to the case where
\begin{equation}
G_{ij, I} =  C_I \frac{ \delta_{ij}}{\left( s^I\right)^2} \quad \quad \text{and} \quad \quad V(s^I,a^I) =V_0 \prod_{I=1}^N \frac{1}{\left( s^I\right)^{\lambda_I}}, \quad \quad \text{with} \quad C_I, \lambda_I >0.
\end{equation}
In terms of canonically normalised fields, this corresponds to exponential saxion potentials for the saxion, with similar terms multiplying the axion kinetic term. Although exponential potentials arise ubiquitously in string theory, we will focus on cases that have a particular relevance for phenomenology or that are particularly robust, as reviewed below.

\subsection{Asymptotic potentials}

 All examples we consider will be compactifications of 10-dimensional type IIB or 11-dimensional F-theory to 4d, where moduli stabilisation is most well developed.\footnote{Which in certain limits are also related to type IIA moduli potentials through mirror symmetry.} A particularly relevant class of potentials are those arising in the so called asymptotic regions of moduli space, where the moduli take large values. This does not only include the regimes of large volume and weak coupling, but more general limits in moduli space involving complex structure moduli in type IIB, such as (but not limited to) the Large Complex Structure (LCS) point. Such asymptotic limits are particularly trustworthy as they allow for parametric control over the computations. From a phenomenological perspective, they are also especially attractive, as they contain ways to generate the hierarchies which appear so abundantly in nature. A classic example of this is given by LVS, where the exponentially large volume naturally induces hierarchies between the susy-breaking, string and Planck scales. More recently, this idea has also been exploited to argue for the existence of a large \emph{Dark Dimension}, motivated by the smallness of the cosmological constant \cite{Montero:2022prj}. In the following, we shall briefly review some of the scenarios to be used in the rest of the paper.

\subsubsection{Large Volume Scenario and variations thereof}\label{ssc:LVS}

A canonical example of asymptotic limit is when the volume of the compactification manifold is taken to be very large, deep into the supergravity regime, providing control over the perturbative $\alpha'$ expansion. The Large Volume Scenario \cite{Balasubramanian:2005zx,Conlon:2005ki} (LVS) is a class of type IIB, $\rm{AdS_4}$ flux vacua with fully broken supersymmetry where all the moduli can be stabilised at exponentially large values of the volume. They are obtained by compactifying type IIB string theory on Calabi-Yau 3-folds with $D3/D7$-branes and $O3/O7$-planes, breaking down the original $\mathcal{N}=2$ supersymmetry to $\mathcal{N}=1$.\footnote{Which is further broken (spontaneously) to $\mathcal{N}=0$ in the vacuum.} The complex structure moduli and the axio-dilaton are stabilised with $F_3$ and $G_3$ flux, as in the GKP construction \cite{Giddings:2001yu}, and acquire masses of order $M_P/\sqrt{\mathcal{V}}$. Therefore, they can be integrated out in the effective description of the K\"ahler moduli $T_i = \tau_i + i b_i$. The latter are a flat direction of the potential at tree level, due to the so-called no-scale structure of the potential, but can be lifted by a delicate balance of perturbative and non-perturbative corrections (to the K\"ahler potential and superpotential respectively). When the latter are included, the $\mathcal{N}=1$ supergravity is specified by
\begin{equation}
K=-2 \log \left(\mathcal{V}+\frac{\xi}{g_s^{3 / 2}}\right) \quad \quad W=W_0+\sum_i A_i e^{-\alpha_i T_i}.
\end{equation}
Aside from the tree-level supergravity term, the K\"ahler potential contains the effect of the $\alpha'^3 \mathcal{R}^4$ supergravity correction, where the coefficient $\xi$ is  given in terms of the Euler characteristic of the Calabi-Yau $\chi(M)$ as $\xi=\frac{\zeta(3) \chi(M)}{2(2 \pi)^3}$. The superpotential is instead a combination of a constant term $W_0$, determined by the values of the stabilised complex structure moduli, and a non perturbative contribution arising from either gaugino condensation or brane instantons. In the simplest case of a Swiss-cheese Calabi-Yau (such as the original example based on the orientifold of $\mathbb{P}_{[1,1,1,6,9]}^4$), the volume can be expressed in terms of a big cycle $T_b$ and small blow-up cycle $T_s$
\begin{equation}\label{eq:vol}
\mathcal{V}=\alpha \left(\tau_b^{3 / 2}-\tau_s^{3 / 2}\right).
\end{equation}
This results in a potential
\begin{equation}\label{eq:Vsb}
V=\frac{A a_s^2 \sqrt{\tau_s} e^{-2 a_s \tau_s}}{\mathcal{V}}-\frac{B W_0 a_s \tau_s e^{-a_s \tau_s}}{\mathcal{V}^2}+\frac{C \xi W_0^2}{g_s^{3 / 2} \mathcal{V}^3},
\end{equation}
where $A,B$ and $C$ are unimportant constants. One can consistently integrate out the blow-up modulus by minimising \eqref{eq:Vsb} with respect to $\tau_s$ and $a_s$, giving
\begin{equation}
V_{\rm{LVS}}=e^{-\frac{\lambda \varphi}{M_P}}\left(A^{\prime}-\varphi^{3 / 2}\right),
\end{equation}
with $\varphi=\sqrt{\frac{2}{3}} \ln \mathcal{V}$ and $\lambda=\sqrt{\frac{27}{2}}$. An interesting variation on the same theme arises when the Calabi-Yau has a fibred structure, and the volume is no longer given by \eqref{eq:vol}. The simplest example is that of a K3 fibration, where the volume can be expressed (in terms of two large K\"ahler moduli $T_1,T_2$) as
\begin{equation}\label{eq:fvol}
\mathcal{V}= \alpha \tau_1 \sqrt{\tau_2}-\gamma \tau_s^{3/2}.
\end{equation}
In exactly the same way as the original LVS, the volume can now be stabilised at exponentially large values. In particular, the potential will take the same functional form as in LVS, but with the volume given by\,\eqref{eq:fvol}
\begin{equation}
V_{LVS_f}(\tau_1,\tau_2)= V_{LVS}(\alpha\tau_1 \sqrt{\tau_2}).
\end{equation}
From the above, it is clear one still has a flat direction corresponding to increasing $\tau_1$ and decreasing $\tau_2$ while keeping the overall volume constant. It can be lifted by string loop correction scaling as $\mathcal{V}^{-10/3}$, and has been used to construct realistic models of inflation\, \cite{Cicoli:2008gp,Cicoli:2011it}.

\subsubsection{Asymptotic potentials in F-theory}\label{ssc:aht}

A much more general class of asymptotic limits involves the complex structure moduli in type IIB compactifications on a Calabi-Yau 3-fold or F-theory compactifications on a 4-fold.\footnote{And eventually their mirror duals.} Using the mathematical tools of asymptotic Hodge theory, it is possible to write down asymptotic expressions for the scalar potentials (and kinetic terms) of moduli approaching the boundary, including in some cases the leading non-perturbative corrections. In the cases of 1- and 2-moduli limits, all possible limits have been classified. Although less developed from the point of view of full moduli stabilisation (K\"ahler moduli are not included in this setting), these scenarios can be derived very rigorously, and are extremely robust in terms of control. Being a vast and complicated subject, we will not attempt to review it here, but only state the most basic results without much motivation. The interested reader can refer to the original papers \cite{Grimm:2019ixq,Bastian:2021eom,Bastian:2021hpc,Grimm:2021ckh,Grimm:2022xmj,Grana:2022dfw,Bastian:2023shf}, as well as \cite{vandeHeisteeg:2022gsp} for an introduction to the topic, \cite{Grimm:2020ouv} for the analysis of axion backreaction in this context and \cite{Calderon-Infante:2022nxb} for cosmological applications.

Such constructions arise from F-theory compactifications with $G_4$ flux on a Calabi-Yau fourfold $Y_4$, with unique, holomorphic $(4,0)$ form $\Omega$. 
Through the duality to M-theory, one can derive the following scalar potential for the complex structure moduli
\begin{equation}
V_M=\frac{1}{\mathcal{V}^3_{Y_4}}\left(\int_{Y_4} G_4 \wedge \star G_4-\int_{Y_4} G_4 \wedge G_4\right),
\end{equation}
where the $G_4$ flux is assumed to be primitive,
\footnote{\emph{I.e.} $J\wedge G_4=0$, with $J$ the K\"ahler form.} 
$\mathcal{V}_{Y_4}$ is the volume and the second term is constrained by the tadpole condition. Equivalently, such a potential can also be derived by an $\mathcal{N}=1$ supergravity with K\"ahler potential and superpotential respectively equal to
\begin{equation}
K^{\mathrm{cs}}=-\log \int_{Y_4} \Omega \wedge \bar{\Omega}, \quad W=\int_{Y_4} \Omega \wedge G_4.
\end{equation}
The crucial point is that, in the asymptotic regions of moduli space, the behaviour of the Hodge norm present in the expressions above simplifies greatly, and can be evaluated (to various degrees of approximation) with the tools of asymptotic Hodge theory. 

If we denote the complex structure moduli as $z^j = s^j+i a^j, j=1...n=h^{3,1}$, we can take $k$ of them to the boundary (while keeping the rest fixed) within a so called growth region, specified by
\begin{equation}
\mathcal{R}_{12 \cdots k}=\left\{\frac{s^1}{s^2}>\gamma, \ldots, \frac{s^{k-1}}{s^k}>\gamma, s^k>\gamma, a^j < \delta; \, \, \text{\rm{for  some}} \,\,\gamma, \delta > 0\right\}.
\end{equation}
The K\"ahler potential then takes the asymptotic form
\begin{equation}\label{eq:kcs}
\begin{split}
K^{\mathrm{cs}} \sim & -\log \left[\left(\frac{s^1}{s^2}\right)^{d_1} \ldots\left(\frac{s^{k-1}}{s^k}\right)^{d_{k-1}}\left(s^k\right)^{d_k} f(z_j,\bar{z}_j)+\ldots\right]=\\
&=-\sum_{j=1}^k \Delta d_j \log s^j-\log f(z_j,\bar{z}_j)+\ldots,
\end{split}
\end{equation}
where $f(z_j,\bar{z}_j)$ denotes the contribution coming from the moduli that are fixed. We have also used $\Delta d_i = d_i-d_{i-1}$ and $d_0=0$, where the $d_i$ are the indices of certain nilpotent matrices associated to the singularities encountered at the points where $s_i \rightarrow \infty$, and they can only take the values $d_i=0...3$. The dots denote both perturbative corrections (in terms of saxion ratios), and non-perturbative corrections with an exponential suppression. Recent advances now allow many of such corrections to be computed reliably, but we shall not be preoccupied with these subtleties and just content ourselves with the leading order expressions. With similar caveats, the scalar potential takes the form
\begin{equation}\label{eq:vcs}
V=\frac{1}{\mathcal{V}_0^3}\left(\sum_{ {{\pmb{\ell}} } \in \mathcal{E}} \prod_{j=1}^{ {\pmb{\ell}} } A_{{\pmb{\ell}}} \left(s^j\right)^{\Delta \ell_j}-A_{\mathrm{loc}}\right).
\end{equation}
The sum is over certain sectors, associated to a splitting of the primitive cohomology $H_{\mathrm{p}}^4\left(Y_4, \mathbb{R}\right)$ near the boundary, specified by an index ${\pmb{\ell}}=(\ell_1,...,\ell_k)$, where $\ell_i=0...8$. As above, we have also define $\Delta \ell_i = \ell_i-\ell_{i-1}$, and $\ell_0=4$. The coefficients $A_{\pmb{\ell}}$ will be polynomials in the axions and the fluxes, which can also depend on the spectator moduli, while $A_{\mathrm{loc}}$ is fixed by the tadpole condition and proportional to the Euler characteristic of $Y_4$. To extract explicit expressions for these coefficients, see for example \cite{Grimm:2019ixq,Calderon-Infante:2022nxb}. From \eqref{eq:kcs}, the leading order metric will be
\begin{equation}
G_{I \bar{J}} = \frac{1}{2}{\rm{Diag}} \left\{ \frac{\Delta d_1}{(s^1)^2},...,\frac{\Delta d_k}{(s^k)^2} \right\}
\end{equation}
The situation we will be interested in is when there is a single leading term in \eqref{eq:vcs} (\emph{i.e} all other terms are suppressed by some positve power of $s$ or $u$), whose coefficient $A_{\pmb{\ell}}$ does not depend on the axions. Since every coefficient $A_{\pmb{\ell}}$ contains a squared flux term $f_i^2$ which does not depend on the axions (while all axion terms are proportional to different fluxes), it is always possible to engineer this condition by turning off some fluxes. In that case, 
the potential can be approximated as
\begin{equation}
V \sim f_i^2  \prod_{j=1}^{ {\pmb{\ell}} } A_{{\pmb{\ell}}} \left(s^j\right)^{\Delta l_j},
\end{equation}
and we will be able to provide analytical results on the corresponding cosmologies, solving the exact equations of motion (and without having to rely on slow roll approximations or similar conditions).

\section{Cosmological evolution}\label{sec:3}
Let us now specialise to the case of a single chiral superfield, and in particular to the limit where it describes a single modulus approaching the boundary of moduli space. Interesting example are the volume (or K\"ahler moduli) in a large volume compactification, or also a complex structure modulus. In either case, the K\"ahler metric at leading order will take the form 
\begin{equation}\label{eq:kls2}
G_{i j} = \frac{C}{ s^2} \delta_{i j},
\end{equation}
where $C$ is an overall numerical constant. For a complex structure modulus $z_i$, as in \eqref{eq:kcs}, $C=\Delta d_i/2$. For solutions where the axion profile is constant, \emph{i.e.} $\dot a=0$, it can always be absorbed by a rescaling of the form $s'=s^{\sqrt{C}}$,\footnote{One must be careful to rescale the potential as well.} but in general it cannot be eliminated. 

With the above metric, the Christoffel symbols for a given saxion-axion pair take the values
\begin{equation}
\Gamma^s_{a\,a}=-\Gamma^a_{s\,a}= -\Gamma^s_{s\,s} = \frac{1}{s} \quad \quad \quad \Gamma^{s}_{s\,a}= \Gamma^{a}_{s\,s}= \Gamma^{a}_{a\,a} =0.
\end{equation}
Overall, the system can be described by the equations of motion for the two fields, supplemented by one of the Friedmann equations. 
\begin{equation}\label{eq:sys1}
\left\{
\begin{aligned}
& \ddot{a}-\frac{2 \dot{a} \dot{s}}{s} + (d-1) H \dot{a}+ \frac{s^2}{C}\partial_a V=0 \\
& \ddot{s}-\frac{\dot{s}^2}{s}+\frac{\dot{a}^2}{s} + (d-1) H \dot{s}+  \frac{s^2}{C} \partial_s V=0\\
&(d-1)(d-2)H^2=  C \frac{\dot{s}^2+\dot{a}^2}{s^2} +2V\\
\end{aligned}\right.
\end{equation}

To describe more general situations, such as cases where $C=0$ and the leading order metric \eqref{eq:kls2} vanishes, one can also consider the more general expression
\begin{equation}\label{eq:kls2gen}
G_{i j} = G(s) \delta_{i j},
\end{equation}
with $G(s)$ some arbitrary function. In this case, the Christoffel symbols take the form
\begin{equation}
\Gamma^s_{a\,a}=-\Gamma^a_{s\,a}= -\Gamma^s_{s\,s} = -\frac{G'(s)}{2 G(s)} \quad \quad \quad \Gamma^{s}_{s\,a}= \Gamma^{a}_{s\,s}= \Gamma^{a}_{a\,a} =0,
\end{equation}
and the corresponding equations of motion become
\begin{equation}\label{eq:sys2}
\left\{
\begin{aligned}
& \ddot{a}+\frac{G'(s)}{G(s)}\dot{a} \dot{s} + (d-1) H \dot{a}+ \frac{\partial_a V}{G(s)}=0 \\
& \ddot{s}+\frac{G'(s)}{2G(s)} \left(\dot{s}^2-\dot{a}^2\right) + (d-1) H \dot{s}+  \frac{\partial_s V}{G(s)} =0\\
& (d-1)(d-2)H^2=  G(s) \left(\dot{s}^2+\dot{a}^2\right) +2V\\
\end{aligned}\right.
\end{equation}
While this form of the equations is not particularly illuminating, it is quite remarkable that for a vanishing axion potential ($ \partial_a V=0$) the first equation can be integrated exactly for any $G(s)$. The result is
\begin{equation}\label{eq:a0g}
\dot{a}(t) = \dot{a}(t_0) \frac{G(s(t_0))}{G(s(t))} e^{- (d-1)\int_{t_0}^{t} d\tau\,  H(\tau)}.
\end{equation}

\subsection{Warm up: axion-saxion kination with no potential}

As a first simplification, let us consider the limit of saxion-axion kination, where the kinetic energy of saxions and axions dominates over the potential $V(s,a)$. This can also describe the case where only the saxion is kinating, and the axion is a flat direction of the potential.  While the kination approximation requires specific initial conditions for the saxions, i.e. that at early times $\frac{1}{2} \dot{s}^2 \gg V(s,a)$,\footnote{Which can be realised by letting the saxion roll along a steep potential after inflation.} it is very natural for the axions. In string theory constructions, axions are often a flat direction of the potential at the perturbative level  - they are protected by a continuous shift symmetry, inherited from higher dimensional gauge invariance. While the latter can be broken to a discrete subgroup by non-perturbative effects, it is a controllable effect in many instances.
We now specialise to \eqref{eq:sys1}, choosing the metric as in \eqref{eq:kls2}. Then\footnote{A similar expression previously appeared in \cite{Calderon-Infante:2022nxb}, without the factor of $s^2(t)$ on the RHS. It was corrected in a subsequent version, after we pointed out Eq. \eqref{eq:a0}.}
\begin{equation}\label{eq:a0}
\dot{a}(t) = \dot{a}(t_0) \frac{s^2(t)}{s^2(t_0)} e^{- (d-1)\int_{t_0}^{t} d\tau\,  H(\tau)}.
\end{equation}
From this equation, one can immediately see that the axion will necessarily evolve during a phase of saxion kination, unless $\dot{a}(t_0)=0$. Assuming a power-law scale factor $ \mathfrak{a}_s(t) \sim t^{\alpha}$, \footnote{To avoid confusion with the axion field $a(t)$, the scale factor in the FLRW metric will be denoted as $\mathfrak{a}_s(t)$ throughout this work.} $H=\frac{\alpha}{t}$, one obtains
\begin{equation}\label{eq:ap1}
\dot{a}(t) = \dot{a}(t_0) \frac{s(t)^2}{s(t_0)^2} \left(\frac{t_0}{t} \right)^{\alpha(d-1)}.
\end{equation}
If $d=4$ and it is the volume that is kinating, $\alpha = 1/(d-1)$ and $s \sim  t^{2/3}$, so that
\begin{equation}\label{eq:ap2}
a(t) =a(t_0)+ \frac{3}{4} \dot{a}(t_0) t_0 \Bigg[\left( \frac{t}{t_0}\right)^{\frac{4}{3}}-1\Bigg]
\end{equation}
and grows with time. While Eqs. \eqref{eq:ap1}-\eqref{eq:ap2} are only valid when $\dot{a}^2 \ll \dot{s}^2$, the equations of motion can actually be solved exactly if the potential (and its derivatives) can be neglected. While setting $V=0$ is not a useful approximation for any realistic scenario, it is nevertheless useful to see how starting from pure saxion kination the axion can also develop some interesting dynamics. From \eqref{eq:sys1}, one can deduce the equation for $H(t)$
\begin{equation}\label{eq:evH}
\dot{H}=-(d-1)H^2,
\end{equation}
which is solved by
\begin{equation}
H(t) = \frac{H_0}{1+H_0(d-1)(t-t_0)}.
\end{equation}
For $t \gg t_0$, this asymptotes to the Hubble rate during kination, $H= \frac{1}{(d-1)t}$. Then, Eq. \eqref{eq:a0} becomes the exact relation
\begin{equation}
\dot{a}(t) = \dot{a}(t_0) \frac{s(t)^2}{s(t_0)^2} \frac{1}{1+H_0(d-1)(t-t_0)},
\end{equation}
and substituting back into \eqref{eq:a0}, this gives an equation for $s$ 
\begin{equation}
\ddot{s}-\frac{\dot{s}^2}{s}+  \frac{\dot{a}^2(t_0)}{s^4(t_0)} \frac{s^3}{\left(1+H_0(d-1)(t-t_0)\right)^2}+\frac{(d-1)H_0 \dot{s}}{1+H_0(d-1)(t-t_0)}=0,
\end{equation}
with the exact solution
\begin{equation}\label{eq:sols}
s(t) =K_1 \frac{s(t_0)^2}{\dot{a}(t_0)} \frac{  H_0(d-1)}{\cosh \left[ K_1\log (1+H_0(d-1)(t-t_0))  -K_2 \right]},
\end{equation}
where 
\begin{equation}
K_1 = \sqrt{\frac{d-2}{d-1}} \quad \quad \quad K_2 = \sinh^{-1} \left( \frac{\dot{s}(t_0)}{\dot{a}(t_0)}\right).
\end{equation}
One can finally integrate \eqref{eq:a0} to obtain
\begin{equation}
a(t)= a(t_0)+ \frac{s(t_0) \dot{s}(t_0)}{\dot{a}(t_0)} \Bigg( 1+\sqrt{1+\left(\frac{\dot{a}(t_0)}{\dot{s}(t_0)}\right)^2}   \tanh \left[ K_1\log (1+H_0(d-1)(t-t_0))-K_2\right] \Bigg).
\end{equation}
At early times, the expressions above simplify to
\begin{equation}
s(t) = s(t_0) \left(\frac{t}{t_0}\right)^{\sqrt{\frac{d-2}{C(d-1)}}}, \quad a(t) = a(t_0)+ \frac{1}{2}\sqrt{\frac{C(d-1)}{d-2}} \dot{a}(t_0) t_0 \Bigg[\left( \frac{t}{t_0}\right)^{2\sqrt{\frac{d-2}{C(d-1)}} }-1\Bigg],
\end{equation}
in accordance with the formulas given above for $d=4$.\footnote{Let us remark that the only four of the initial conditions given by  $\{s(t_0),\dot{s}(t_0),$ $a(t_0),\dot{a}(t_0),H_0\}$ are independent, since they are related by the constraint equation \eqref{eq:ap2} $(d-1)(d-2)H_0= C \frac{\dot{s}^2(t_0)+\dot{a}^2(t_0)}{s^2(t_0)}.$ }

\subsection{Including a potential for the saxions}\label{sc:gs}

In a realistic scenario, the presence of a  potential will spoil the pure kination approximation used above, when either $\dot{a}^2$ or $\dot{s}^2$ fall below $V$. Let us then study the system \eqref{eq:sys1}, where we have a potential $V(s_i)$ for the saxions which can also induce interactions between them. For simplicity, we will for now keep the axion as a exact flat direction, so that $\partial_{a_i} V=0$. This is an excellent approximation for K\"ahler axions paired to saxions parametrising large cycles in a compactification (for example the volume in LVS), as their potential can only come form non-perturbative terms which are highly suppressed. In the case of asymptotic, complex structure moduli, although the dominant saxion term is always independent of the axions, there can be polynomials in $\frac{a}{s}$ multiplying the leading order potential, which can become important if $a \sim s$. Such additional terms can be turned off by a judicious choice of fluxes, and that is what will be assumed in this work. \footnote{Although they could, in principle, be treated in this framework, the problem becomes highly specific, and it is left to future work.}

\subsubsection{General solution}\label{ssc:gen}

Let us consider the equations of motion \eqref{eq:eom}-\eqref{eq:f1} for $N$ chiral multiples $\Phi^I$. The equivalent of \eqref{eq:evH} can no longer be integrated directly, as
\begin{equation}\label{eq:Hff}
\dot{H} = -(d-1)H^2+\frac{2 V(s_i)}{d-2},
\end{equation}
and we will adapt a different approach. In analogy with the standard treatment of cosmological tracker solutions \cite{Copeland:1997et,Ferreira:1997hj}, we can introduce the new variables
\begin{equation}\label{eq:xiyi}
x_i = \sqrt{\frac{C_i}{(d-1)(d-2)}} \frac{\dot{s_i}}{H s_i} \quad \quad y_i = \sqrt{\frac{C_i}{(d-1)(d-2)}} \frac{\dot{a_i}}{H s_i} \quad \quad z =\frac{1}{H} \sqrt{\frac{2V(s)}{(d-1)(d-2)}},
\end{equation}
where $i$ ranges from $1$ to $N$, and which satisfy
\begin{equation}
\sum_{j=1}^N(x_j^2+y_j^2)+z^2 =1.
\end{equation}
Incidentally, \eqref{eq:Hff} allows one to write the acceleration parameter as
\begin{equation}
\varepsilon \equiv - \frac{\dot{H}}{H^2} = (d-1)\left(1-z^2\right).
\end{equation}
Therefore, in terms of these new variables, accelerated expansion can only be achieved when
\begin{equation}
\sum_{i=1}^N x_i^2+y_i^2< \frac{1}{d-1}.
\end{equation}
The equations of motion can also be written in the more transparent form
\begin{equation}\label{eq:sysN}
\left\{
\begin{aligned}
\frac{d x_i}{dM}&=- \sqrt{\frac{(d-1)(d-2)}{C_i}} y_i^2 -\left(1-\sum_{j=1}^N (x_j^2+y_j^2)\right) \left[(d-1)x_i+\frac{s_i \partial_i V}{2V} \sqrt{\frac{(d-1)(d-2)}{C_i}} \right]\\
\frac{d y_i}{dM}&=  \sqrt{\frac{(d-1)(d-2)}{C_i}}  x_i y_i-(d-1)\left(1-\sum_{j=1}^N (x_j^2+y_j^2)\right)y_i
\end{aligned}\right.,
\end{equation}
where the new time coordinate is given in terms of the number of e-folds  $M= \log \mathfrak{a}_s$, with $\mathfrak{a}_s$ the scale factor (not to be confused with an axion). As in the case of a single field, additional assumptions are needed in order to turn this into an autonomous system in term of the $\{x_i,y_i\}$ variables only. In particular, we will take $V(s_i)$ to be of the product form
\begin{equation}\label{eq:potf}
V(s_i) = V_0 \prod_{j=1}^N s_j^{-\lambda_j},
\end{equation}
such that the combination appearing in \eqref{eq:sysN} (where indices are not summed over) gives
\begin{equation}\label{eq:potlambda}
\frac{ s_i \partial_i V}{V} = - \lambda_i.
\end{equation}
This is motivated from the asymptotic form of scalar field potentials close to the boundary of moduli space (as discussed in Section \ref{sec:mot}), where the leading terms are always monomials in the saxions and independent from the axions. \footnote{If one found critical points satisfying $y_i \ll x_i$, they could be valid asymptotic solutions even if the potential \eqref{eq:potlambda} contained additional terms scaling with positive powers of $\frac{a_i}{s_i}$.} While the choice of a single, dominating term in the potential might appear particularly restrictive, there are physically interesting situations where indeed this is the case. An example is when the potential is obtained as an expansion in a single large quantity which can be expressed as the product of two or more moduli (such as the compactification volume in a fibred Calabi-Yau, see Subsection \ref{ssc:LVm}). Then, the leading term in the potential will be of the form \eqref{eq:potlambda}. In more general asymptotic limits, certain terms in the potential can in principle be of the same order along the asymptotic trajectory.\footnote{Notice that terms where both (all) moduli appear with a higher power are automatically subleading.} However, in many cases of interest they can be turned off by setting specific fluxes to zero, or made negligible with a sufficient flux hierarchy. We expect the case with competing terms in the potential to be significantly more complicated, as the same system without axions already proves challenging to analyse analytically \cite{Collinucci:2004iw}.

Let us now analyse the general solution to \eqref{eq:sysN}, assuming also \eqref{eq:potlambda} holds. A first thing to notice is that \eqref{eq:sysN} is invariant under $y_i \rightarrow -y_i$, and the lines $y_i=0$ correspond to invariant manifolds (they are preserved under time evolution, and cannot be crossed by any trajectory). Without loss of generality, we can then assume $y_i \geq 0$. 

The fixed points of \eqref{eq:sysN} have to satisfy $\frac{dx_i}{dM}\equiv \frac{dy_i}{dM}=0$ for every $i$, with the condition that $0 \leq z^2 \leq 1$. For the latter equations, a trivial solution is the one where $y_i=0$. Therefore, for a given solution it is always possible to reshuffle the $y_i$ in such a way that
\begin{equation}
y_i \neq 0 \quad \text{for} \quad i=1...K \quad \quad \text{and} \quad  \quad y_i =0 \quad \text{for} \quad i=K+1...N.
\end{equation}
For now, let us also assume that $z\neq 0$. It follows that the fixed points (denoted with a bar in the rest of this section) are given by
\begin{equation}\label{eq:fp1}
\bar{x}_i= \sqrt{\frac{C_i(d-1)}{d-2}} \bar{z}^2 \quad \quad \quad
\bar{y}_i=\bar{z}^2 \sqrt{  \frac{\lambda_i}{2 \bar{z}^2}-\frac{C_i(d-1)}{d-2}}
 \quad \text{for} \quad i \leq K, 
\end{equation}
and
\begin{equation}\label{eq:fp2} 
\bar{x}_i=\frac{\lambda_i}{2}\sqrt{\frac{d-2}{C_i(d-1)}} \quad \quad \quad \quad \quad \quad  \bar{y}_i=0  \quad \quad \quad \quad \quad \quad\text{for} \quad i>K,
\end{equation}
where
\begin{equation}\label{eq:zz}
\bar{z}^2 =1-\sum_{j=1}^N (\bar{x}_j^2+\bar{y}_j^2) =\frac{2- \frac{d-2}{2(d-1)} \sum_{j=K+1}^N \frac{\lambda_j^2}{C_j} }{2 +\sum_{j=1}^K \lambda_j}.
\end{equation}
From \eqref{eq:fp1} and \eqref{eq:zz}, a valid solution exists as long as
\begin{equation}\label{eq:cond1}
 4  \frac{d-1}{d-2}-\frac{\lambda_i}{C_i}(2+\sum_{j=1}^K \lambda_j) \leq  \sum_{j=K+1}^N \frac{\lambda_j^2}{C_j} \leq 4 \frac{d-1}{d-2}, \quad \quad \text{for any} \quad i \leq K.
\end{equation}
As a particular case, if $\lambda_i^2 >4 C_i \frac{d-1}{d-2} $ for all $i$, there are no solutions with $K<N$. In that case, the only non-trivial fixed point is
\begin{equation}\label{eq:fpp}
\begin{split}
\bar{x}_i&=\frac{2}{2+\sum_{j=1}^N \lambda_j} \sqrt{\frac{C_i(d-1)}{d-2}}  \\ \bar{y}_i&= \frac{2}{2+\sum_{j=1}^N \lambda_j} \sqrt{\frac{C_i(d-1)}{d-2}} \, \sqrt{\frac{\lambda_i(2+\sum_{j=1}^N \lambda_j)}{4}\frac{d-2}{C_i(d-1)}-1},
\end{split}
\end{equation}
and
\begin{equation}
\bar{z}^2=1-\sum_{j=1}^N (\bar{x}_j^2+\bar{y}_j^2)=\frac{2}{2+\sum_{j=1}^N \lambda_j}.
\end{equation}

Coming back to the possibility $z=0$, in that case there exists an invariant manifold given by
\begin{equation} \label{eq:inma}
\sum_{i=1}^N \bar{x}_i^2 =1 \quad \quad \quad \quad \bar{y}_i= 0 \quad \quad \quad \quad \text{for} \quad\quad i=1...N,
\end{equation}
corresponding to kination along an arbitrary direction in saxion space. However, any critical point lying on this surface is always unstable. To see this, one can evaluate the Jacobian $\mathcal{J}$ of the system \eqref{eq:sysN} at any point of the type \eqref{eq:inma}, which takes the block form
\begin{equation}
\mathcal{J} \lvert_{x_i=\bar{x}_i, y_i=0}= 
\begin{pmatrix}
J_X  & 0\\
0 &J_Y \\
\end{pmatrix}.
\end{equation}
In turn, the two $N \times N$ blocks are defined by
\begin{equation}
\left(J_X\right)_{i j}= -2 \bar{x}_j \left( (d-1)\bar{x}_i-\frac{\lambda_i}{2}\sqrt{\frac{(d-1)(d-2)}{C_i}} \right)  \quad \quad \left(J_Y\right)_{i j} = \delta_{ij} \sqrt{\frac{(d-1)(d-2)}{C_i}} \bar{x}_i.
\end{equation}
If any of the $\bar{x}_i$ are positive, $J_Y$ has a positive eigenvalue. If they are all negative, however, $J_X$ has only positive elements, and must have at least one positive eigenvalue. Therefore, stability is never achieved.

To determine if any of the other solutions are stable attractors, one can perform a linear stability analysis exactly as we have done above. For the special cases $N=1,2$ this has been carried out in appendix \ref{sec:app}. The general case requires a bit more caution, and this will be the topic of the next section. Before we do that, however, we shall use some results from stability theory to gain some insight into the behaviour of the system (See \cite{Wiggins:2003} for a review).

In the region defined by  $x_i \geq \frac{\lambda_i}{2}\sqrt{\frac{d-2}{C_i(d-1)}} $, $\frac{d x_i}{dM}$ is semi-negative definite from  \eqref{eq:sysN}. Using $x_i(M)$ as a Lyapunov function, and by applying \emph{La Salle's} invariance principle, any trajectory that is entirely contained in this subspace will asymptotically converge to the locus where $ \frac{d x_i}{dM}=0$ , that is $y_i=0$ and $x_i = \frac{\lambda_i}{2}\sqrt{\frac{d-2}{C_i(d-1)}}$. On the other hand, if at any point of the evolution $x_i(M_0) < \frac{\lambda_i}{2}\sqrt{\frac{d-2}{C_i(d-1)}}$, it follows from the equation of motion for $x_i$ that $x_i(M) \leq \frac{\lambda_i}{2}\sqrt{\frac{d-2}{C_i(d-1)}}$ for any $M>M_0$.\footnote{When $x_i = \frac{\lambda_i}{2}\sqrt{\frac{d-2}{C_i(d-1)}}$, $\frac{d x_i}{dM} <0$.} Overall, we conclude the set $x_i\leq \frac{\lambda_i}{2}\sqrt{\frac{d-2}{C_i(d-1)}}$ is a global attractor set for the system under consideration, which is a non-trivial statement whenever $\frac{\lambda_i}{2}\sqrt{\frac{d-2}{C_i(d-1)}} < 1$. From this, one might be tempted to conclude that if
\begin{equation}\label{eq:cond}
\lambda_i<2 C_i \frac{d-1}{ d-2} \left( 1-\sum_{j=1}^N x_i^2+y_i^2\right)\Bigg \rvert_{x_i= \bar{x}_i,{y_i= \bar{y}_i} } 
\end{equation}
the fixed point \eqref{eq:fp1} (with $i<K$) cannot be stable, and \eqref{eq:fp2} is the only possibility. However, notice that Eq. \eqref{eq:cond} is the exactly the complement of the first condition in \eqref{eq:cond1}, so we have already shown that when it is satisfied the fixed point \eqref{eq:fp1} cannot even exist (which is a stronger statement).

As a next step, it is instructive to define the \emph{Lyapunov} function
\begin{equation}
\mathbf{L}(x_i,y_i)= \sum_{j=K+1}^N \left( x_i-\frac{\lambda_i}{2}\sqrt{\frac{d-2}{C_i(d-1)}} \,\right)^2+y_i^2,
\end{equation}
which is positive definite on the whole domain and vanishes only if all the terms in the sum are individually zero. Taking a derivative with respect to $M$,
\begin{equation}\label{eq:lyp}
\begin{split}
\frac{d\, \mathbf{L}(x_i(M),y_i(M))}{dM} = &-2 (d-1) \left[ 1-\sum_{j=1}^N (x_j^2+y_j^2)\right] \sum_{j=K+1}^N    \left( x_i-\frac{\lambda_i}{2}\sqrt{\frac{d-2}{C_i(d-1)}} \,\right)^2 \\&-
\sum_{j=K+1}^N 
\left( 2(d-1)\left[ 1-\sum_{j=1}^N (x_j^2+y_j^2)\right] -(d-2)\frac{\lambda_i}{C_i}\right)y_i^2.
\end{split}
\end{equation}
If, at the fixed point, the $\lambda_i$ all satisfy
\begin{equation}\label{eq:condl}
\lambda_i<2 C_i \frac{d-1}{ d-2} \left( 1-\sum_{j=1}^N x_i^2+y_i^2\right)\Bigg \rvert_{x_i= \bar{x}_i,{y_i= \bar{y}_i} } \quad \quad i=K+1...N,
\end{equation}
one can define the non-empty set containing $\bar{x}_i,\bar{y}_i$,
\begin{equation}
\mathcal{S}_A =\left\{ x_i,y_i \quad | \quad \lambda_i<2 C_i \frac{d-1}{ d-2} \left( 1-\sum_{j=1}^N  x_i^2+y_i^2\right)  \quad \quad i=K+1...N \right\}, 
\end{equation}
and inside which $\mathbf{L}$ is monotonically decreasing. By \emph{La Salle's} invariance principle, all trajectories that remain within $\mathcal{S}_A$ will converge asymptotically to the set $\mathcal{S}_B$ defined as
\begin{equation}
\mathcal{S}_B=
\left\{ x_i=\frac{\lambda_i}{2}\sqrt{\frac{d-2}{C_i(d-1)}} , y_i=0\right\}_{ i=K+1...N}.
\end{equation}
It is still not clear from the above that all the trajectories in a neighbourhood of the fixed point will remain within the set $\mathcal{S}_B$,
and therefore that \eqref{eq:condl} guarantees stability. However,  \eqref{eq:condl} is exactly the same condition that will come up when analysing the perturbative stability about the fixed points. This is perhaps hinting towards the existence of a suitable modification of \eqref{eq:lyp} serving as a global \emph{Lyapunov} function for the system, although we were not able to find one.

To determine the stability of a fixed point of the form given in \eqref{eq:fp1}-\eqref{eq:fp2}, we can consider the following auxiliary system, in one dimension extra,
\begin{equation}\label{eq:sysN2}
\left\{
\begin{aligned}
\frac{d S}{dM}&=-2(1-S) \left[ (d-1)S- \sum_{i=1}^N \frac{\lambda_i \alpha_i}{2}x_i \right]\\
\frac{d w_i}{dM}&=  2  w_i \big[ \alpha_i x_i-(d-1)(1-S) \big] \quad \quad\quad \quad \quad \quad i=1...K\\
\frac{d x_i}{dM}&=- \alpha_i  w_i -\left(1-S\right) \left[(d-1)x_i-\frac{\lambda_i \alpha_i}{2}  \right] \quad \quad i=1...K\\
\frac{d w_i}{dM}&=  2  w_i \big[ \alpha_i x_i-(d-1)(1-S) \big] \quad \quad\quad \quad \quad \quad i=K+1...N\\
\frac{d x_i}{dM}&=- \alpha_i  w_i -\left(1-S\right) \left[(d-1)x_i-\frac{\lambda_i \alpha_i}{2}  \right] \quad \quad i=K+1...N\\
\end{aligned}\right.
\end{equation}
where we have defined
\begin{equation}
w_i = y_i^2 \quad \quad \text{and} \quad  \quad \alpha_i= \sqrt{\frac{(d-1)(d-2)}{C_i}},
\end{equation}
and the other variables coincide with the ones used before. We can then consider the fixed point given by
\begin{equation}\label{eq:fp3}
\bar{S}=\sum_{i=1}^N \frac{\lambda_i \alpha_i}{2(d-1)}\bar{x}_i,
\end{equation}
with
\begin{equation}\label{eq:fp4}
\bar{x}_i \equiv \frac{(d-1)(1-\bar{S})}{\alpha_i}  \quad \quad \bar{w}_i \equiv \frac{\lambda_i}{2} (1-\bar{S}) - \frac{C_i(d-1)}{d-2}(1-\bar{S})^2 \quad \quad i \leq K
\end{equation}
and
\begin{equation}\label{eq:fp5}
\,\bar{x}_i \equiv \frac{\lambda_i \alpha_i}{2(d-1)}  \quad \quad \quad \quad \quad \quad \quad \quad \quad \quad \bar{w}_i \equiv 0 \quad \quad \quad \quad \quad \quad \quad  i > K.
\end{equation}
From these definitions, \eqref{eq:sysN2} is equivalent to \eqref{eq:sysN} on the hypersurface $\mathcal{S}_N$ defined by $S= \sum_{i=1}^N x_i^2 +y_i^2$. With this identification, the fixed points in \eqref{eq:fp3}-\eqref{eq:fp5} also map to those in Eqs. \eqref{eq:fp1} and \eqref{eq:fp2}. By construction, $\mathcal{S}_N$ contains all trajectories intersecting it, since
\begin{equation}
\frac{d \left( S-  \sum_{i=1}^N x_i^2 +y_i^2\right)}{dM} =0 \quad \quad \quad \text{for} \quad \quad \quad S-  \sum_{i=1}^N x_i^2 +y_i^2=0,
\end{equation}
and is an invariant manifold for \eqref{eq:sysN2}. Therefore, stability of the fixed points of the form \eqref{eq:fp3}-\eqref{eq:fp5} for the system \eqref{eq:sysN2} implies the stability of the corresponding points \eqref{eq:fp1}-\eqref{eq:fp1} for the system \eqref{eq:sysN}. Notice, however, that the \emph{instability} of \eqref{eq:sysN2} does \emph{not} necessarily imply that of \eqref{eq:sysN} , as it could be that only some trajectories outside the hypersurface are repelled from the fixed point. From a local stability perspective, the latter can happen only if the Jacobian of \eqref{eq:sysN2} at a fixed point has exactly one eigenvalue $\kappa_1$ with negative real part  (with eigenvector $v_1$). In that case, the corresponding fixed point for \eqref{eq:sysN} is stable if and only if the other eigenvectors $\left\{ v_2...v_n \right\}$ all lie on the tangent plane to $\mathcal{S}_N$.

The advantage of this reformulation lies in the fact that the Jacobian $\mathcal{J}$ of the matrix  defining the system \eqref{eq:sysN2}, evaluated at the fixed point \eqref{eq:fpp}, takes a simpler form, since the coordinates $x_i,y_i$ no longer mix with each other, but only with the new variable $S$. In particular, said Jacobian can be expressed as the block matrix 
\begin{equation}
\mathcal{J} \lvert_{x_i = \bar{x}_i, w_i = \bar{w}_i,S=\bar{S}_i}= 
\begin{pmatrix}
J_1  & J_3\\
0 &J_2 \\
\end{pmatrix}
\end{equation}
where the coordinates are ordered as in \eqref{eq:sysN2} and the blocks $J_1$ and $J_2$ have dimensions $(2K+1) \times (2K+1)$ and $2(N-K) \times 2(N-K)$ respectively. Therefore, the eigenvalues of $\mathcal{J}$ are the same as the eigenvalues of $J_1$ and $J_2$, which we now analyse.

The first block takes the following form:
\begin{equation}
J_1=\begin{pmatrix}
 A  & \begin{matrix} 0 & B_1 & \dots & 0 & B_K \end{matrix} \\
 \begin{matrix} C_1  \\ D_1 \\ \vdots \\[1ex] C_K \\D_K \end{matrix} &
  \text{\Huge$M$}
\end{pmatrix}.
\end{equation}
The elements on the first row/column are defined as
\begin{equation}\label{eq:el1}
A= -2(d-1)(1-\bar{S}) \quad B_i= \lambda_i \alpha_i (1-\bar{S}) 
\quad C_i= 2(d-1) \bar{w_i} \quad D_i = -\frac{\alpha_i \bar{w_i}}{1-\bar{S}},
\end{equation}
and in turn the sub-matrix $M$ takes the form
\begin{equation}
M= \begin{pmatrix}
  \begin{matrix}
  0 & E_1 \\
  F_1 & G_1
  \end{matrix}
  & \rvline & \bigzero 
  & \rvline & 
  \begin{matrix}
  \hdots & \hdots \\
  \hdots & \hdots
  \end{matrix}
  & \rvline & \bigzero  \\
  
\hline
  \bigzero & \rvline &
  \begin{matrix}
  0 & E_2 \\
  F_2 & G_2
  \end{matrix}
  & \rvline & \begin{matrix}
  \hdots & \hdots \\
  \hdots & \hdots
  \end{matrix}
  & \rvline & \bigzero  \\
  
  \hline
  
  \bigzero & \rvline &
  \bigzero & \rvline & 
  \begin{matrix}
  \hdots & \hdots \\
  \hdots & \hdots
  \end{matrix}  & \rvline &  
     \bigzero  \\
  
  \hline
  \bigzero & \rvline &
  \bigzero & \rvline &
   \begin{matrix}
  \hdots & \hdots \\
  \hdots & \hdots
  \end{matrix}  & \rvline  
  &  \begin{matrix}
  0 & E_N \\
  F_N & G_N
  \end{matrix}   \\
  \end{pmatrix},
\end{equation}
with
\begin{equation}\label{eq:el2}
E_i = 2 \alpha_i \bar{w_i} \quad F_i= - \alpha_i \quad G_i=-(1-S)(d-1).
\end{equation}
Thanks to the block form of $M$, the characteristic polynomial can be written as
\begin{equation}\label{eq:chpo}
\text{det} (J- \kappa \mathbb{1}) = A \prod_{i=1}^N  \text{det} \left[ \begin{pmatrix}
  -\kappa & E_i \\
  F_i & G_i-\kappa
  \end{pmatrix} \right] + \sum_{j=2}^N B_j(C_j F_j+\kappa D_j) \prod_{i=1 , i\neq j}^N  \text{det} \left[\begin{pmatrix}
  -\kappa & E_i \\
  F_i & G_i-\kappa
  \end{pmatrix} \right],
\end{equation}
Substituting the values \eqref{eq:el1}-\eqref{eq:el2} in \eqref{eq:chpo}, and dividing by a non-vanishing common factor, one obtains the following eigenvalue equation
\begin{equation}\label{eq:eig}
2(d-1)(1-\bar{S}) + \left(\kappa + 2(d-1)(1-\bar{S})\right) \sum_{j=2}^N \frac{\lambda_j \alpha_j^2 \bar{w_j}}{\lambda^2+\lambda (1-\bar{S)}(d-1)+2 \alpha_j^2 \bar{w_j}}=0.
\end{equation}
Although the solutions to \eqref{eq:eig} cannot be determined analytically, we are only concerned with the sign of the real part of $\kappa$. If we write $\kappa=p+ i q$, and consider both the real and imaginary part of the above equation, we can re-express it as
\begin{equation}
\left( 2p + (d-1)(1-\bar{S}) \right) \sum_{j=2}^N \frac{\lambda_j \alpha_j^2 \bar{w_j}}{|\lambda^2+\lambda (1-\bar{S)}(d-1)+2 \alpha_j^2 \bar{w_j}|^2} = -\frac{2(d-1)(1-\bar{S})}{ \left(p+2(d-1)(1-\bar{S}) \right)^2}.
\end{equation}
Since the sum on the LHS is positive, this implies 
\begin{equation}
p < -\frac{(d-1)(1-\bar{S)}}{2},
\end{equation}
proving that every eigenvalue of $J_1$ has a negative real part, corresponding to a stable solution.

The second block takes the much simpler, diagonal form
\begin{equation}
J_2= 
\begin{pmatrix}
 \text{Diag} \left(2  \left[\alpha_{K+i} \bar{x}_{K+i}-(d-1)(1-\bar{S}) \right] \right)_{i=1...N-K}\\
0 & -(d-1)(1-\bar{S})  \mathbb{1}_{N-K}
\end{pmatrix}
\end{equation}
In this case, it is immediate to extract the eigenvalues, and to conclude that the system \eqref{eq:sysN2} is stable if and only if 
\begin{equation}\label{eq:stag}
\alpha_{i} \bar{x}_{i}<(d-1)(1-\bar{S}) \quad \quad \quad \quad i > K.
\end{equation}
Substituting Eqs. \eqref{eq:fp4}-\eqref{eq:fp5} into the above, this is equivalent to the condition \eqref{eq:condl} being satisfied, confirming the intuition gained using the Lyapunov functions. Therefore, we conclude that a fixed point of the form \eqref{eq:fp1}-\eqref{eq:fp2} is an attractor if it satisfies the existence conditions in \eqref{eq:cond} and
\begin{equation}\label{eq:conds}
 4  \frac{d-1}{d-2}-\frac{\lambda_i}{C_i}(2+\sum_{j=1}^K \lambda_j) > \sum_{j=K+1}^N \frac{\lambda_j^2}{C_j}\quad \quad \text{for} \quad i > K.
\end{equation}
are satisfied. Since some of the other eigenvectors of $J_2$ do not lie on the tangent plane to $\mathcal{S}_N$ at the critical point, \eqref{eq:conds} is both a necessary and sufficient condition for stability, even for the system \eqref{eq:sysN}. Notice that \eqref{eq:conds} is the exact complement of the conditions that the $\lambda_i, C_i$ with $i \leq K$ need to satisfy for the fixed point to exist.

\subsubsection{Single modulus}

For a single field, the perturbative stability analysis of the fixed points can be easily carried out by computing the eigenvalues of the Hessian, as was already carried out in \cite{Sonner:2006yn,Russo:2018akp,Russo:2022pgo} (and in \cite{Cicoli:2020cfj,Cicoli:2020noz,Brinkmann:2022oxy,Christodoulidis:2021vye} in the presence of matter, see the next section). A detailed analysis is presented in Table \ref{tab:om}, and we refer to Appendix \ref{app:a1} for a more general treatment. The outcome is that, for a fixed value of $\lambda$, there is always a single dynamical attractor for the system. The trivial critical points $\bar{x}=\pm 1, \bar{y}=0$ correspond to saxion kination, but they are never an attractor. For $\lambda < \sqrt{1+\frac{4C(d-1)}{d-2}}-1$, this is given by\footnote{This fixed point exists for any $\lambda \leq 2 \sqrt{\frac{C(d-1)}{d-2}}$, but is not stable for $\lambda > \sqrt{1+\frac{4C(d-1)}{d-2}}-1$.}
\begin{equation}
\bar{x} = \frac{\lambda}{2}\sqrt{\frac{d-2}{C(d-1)}} \quad \quad  \bar{y}=0 \quad \quad \bar{z}^2=1-\frac{(d-2) \lambda ^2}{4 (d-1) C}.
\end{equation}
This solution is well known, and corresponds to a scaling solution along an exponential potential for the saxion, with a power law scale factor. Indeed, it is characterised by
\begin{equation}
H(t) = \frac{4C}{(d-2) \lambda^2} \frac{1}{t} 
\end{equation}
and
\begin{equation}
\quad \quad s(t) =\left[ \frac{\lambda^4 (d-2)V_0}{2C} \frac{1}{4C(d-1)-(d-2) \lambda^2} \right]^{1/\lambda} t^{2/\lambda}  \quad \quad \quad a(t) = a(t_0).
\end{equation}
However, for any $\lambda > \sqrt{1+\frac{4C(d-1)}{d-2}}-1$, there is also a new fixed point, thanks to the kinetic term coupling the axion with the saxion. It is given by
\begin{equation}\label{eq:fp}
\bar{x}= \frac{2}{2+\lambda} \sqrt{\frac{C(d-1)}{d-2}} \quad \quad \quad \bar{y}= \frac{2}{2+\lambda}  \sqrt{\frac{C(d-1)}{d-2}}  \sqrt{\frac{\lambda(\lambda+2)}{4} \frac{d-2}{C(d-1)}-1},
\end{equation}
and it is always stable (whenever it exists). It is always consistent with the (Friedmann) constraint equation, as
\begin{equation}
\bar{z}^2=1-\bar{x}^2-\bar{y}^2=\frac{2}{\lambda+2}.
\end{equation}
By inserting a power law ansatz for $s(t)$ and $a(t)$, one can verify that the attractor solution takes the form
\begin{equation}
s(t) = \left( \frac{\lambda^2}{\lambda+2} \frac{d-1}{d-2}\right)^{1/\lambda} t^{2/\lambda} \quad \quad a(t) = \sqrt{\frac{\lambda(\lambda+2)}{4} \frac{d-2}{d-1}-1}\left( \frac{\lambda^2}{\lambda+2} \frac{d-1}{d-2}\right)^{1/\lambda} t^{2/\lambda}
\end{equation}
and
\begin{equation}\label{eq:HA}
H(t) = \frac{2+\lambda}{\lambda(d-1)t}.
\end{equation}
Indeed, evaluating $x$ and $y$ for the above solutions one obtains exactly the values in \eqref{eq:fp}. For large $\lambda$, this reduces to the asymptotic solution in the case of pure kination found above. 

\begin{table}[ht]
\begin{center}
\begin{tabular}{ | c | c | c | c |  }
    \hline
    $x$ & $y$  & Existence & Stability \\ \hline \hline
    $\pm 1$ & $0$  & always & No \\ \hline
    $\frac{\lambda}{2} \sqrt{\frac{C(d-1)}{d-2}}$ & $0$  & $\lambda^2 < 4\frac{C(d-1)}{d-2} $ & $\lambda(\lambda+2) < 4\frac{C(d-1)}{d-2} $ \\
    \hline
    $\frac{2}{2+\lambda} \sqrt{\frac{C(d-1)}{d-2}}$ &  $\frac{2}{2+\lambda}    \sqrt{\frac{\lambda(\lambda+2)}{4} - \frac{C(d-1)}{d-2}}$  & $\lambda(\lambda+2) > 4 C \frac{d-1}{d-2} $ & always \\
    \hline
  \end{tabular}
  \caption{Critical points of the system \eqref{eq:sysN}, for a single modulus ($N=1$). The stability column assumes the existence conditions are verified. More details are presented in Appendix \ref{app:a1}.}
  \label{tab:om}
  \end{center}
\end{table}

\subsubsection{Two moduli}\label{ssc:tm}

Already in the case of two moduli, the brute force computation of the eigenvalues of \eqref{eq:sysN} becomes very involved, and it is not possible to obtain analytic expressions for all the critical points (some partial results are provided in Appendix \ref{app:1b}). Therefore, the more general techniques introduced above are already necessary to study the stability of the system. Let us just enumerate the stable, critical points, and refer to Appendix \ref{app:1b} for a more detailed discussion.

\begin{figure}[h!]
    \centering
    \includegraphics[width=1.0\textwidth]{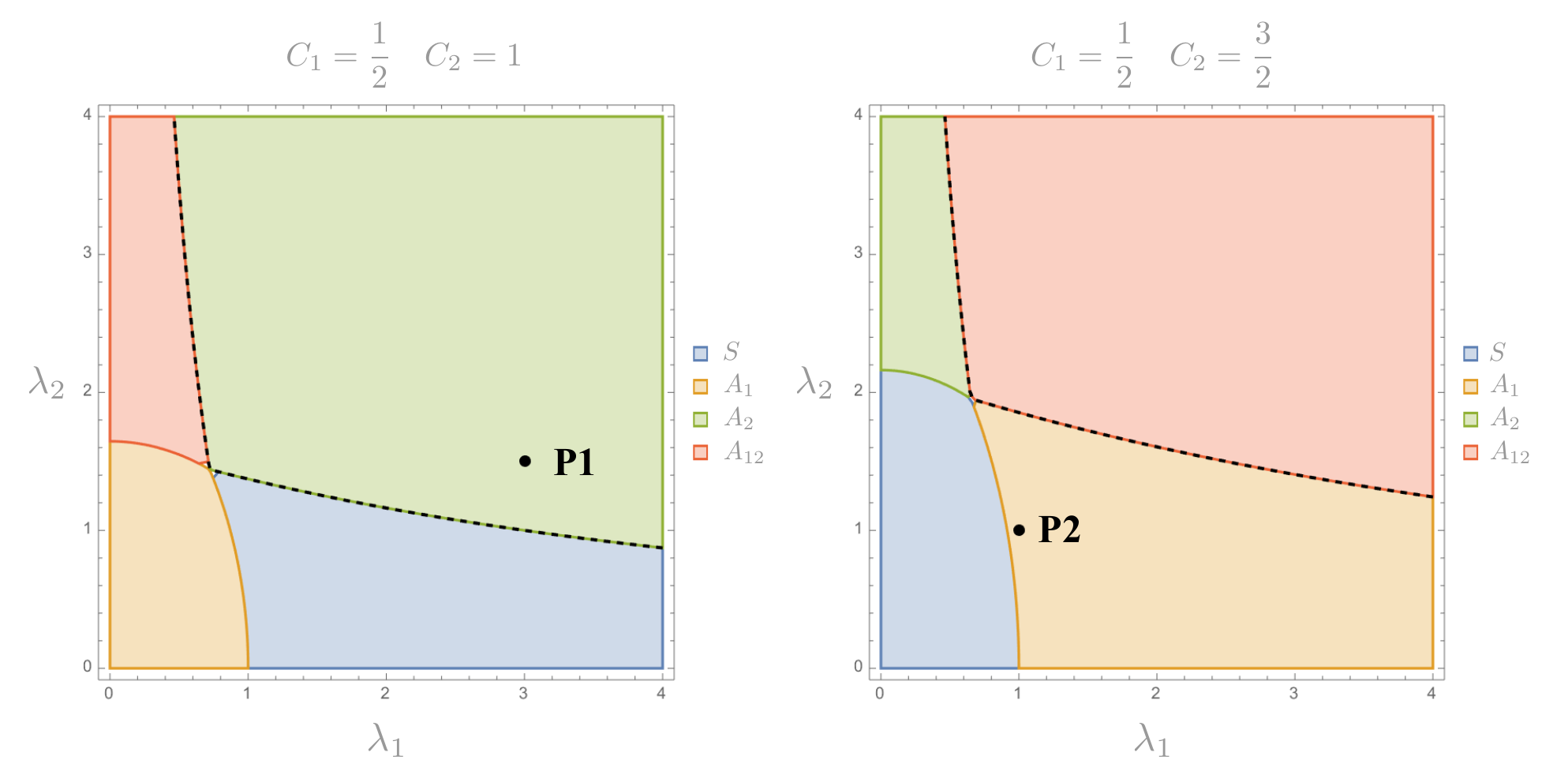}
    \caption{Diagram depicting the regions of stability for the system \eqref{eq:sysN} with two moduli, for fixed values of $C_1$ and $C_2$. Shaded regions indicate the existence of a stable, critical point, denoted with the same notation employed in the main text. The coloured regions result from a numerical calculation of the eigenvalues, and they match with the analytical conditions \eqref{eq:2ms}-\eqref{eq:axfp} (denoted by dotted lines). The points P1 and P2 are two benchmark points discussed in Section \ref{sec:app}, given by \eqref{eq:LVSf} and \eqref{eq:solax} respectively.}
    \label{fig:l12}
\end{figure}

The first critical point is the one where the axion are constants, namely
\begin{equation}
\bar{x}_1= \frac{\lambda_1}{2} \sqrt{\frac{d-2}{C_1(d-1)}} \quad \quad \bar{x}_2= \frac{\lambda_2}{2} \sqrt{\frac{d-2}{C_2(d-1)}} \quad \quad \bar{y}_1=\bar{y}_2=0,
\end{equation}
and we denote it as $S$. From a direct computation of the eigenvalues, it exists and is stable for
\begin{equation}
\frac{\lambda_1^2}{C_1}+\frac{\lambda_2^2}{C_2} < {\rm{Min}} \left\{ 4\frac{d-1}{d-2}-\frac{2 \lambda_1}{C_1}, 4\frac{d-1}{d-2}-\frac{2\lambda_2}{C_2} \right\}.
\end{equation}
The critical point with one constant axion, denoted as $A_1$ is instead given by\footnote{There is of course an analogous point $A_2$ obtained by the permutation $\lambda_1 \leftrightarrow \lambda_2$.}
\begin{equation}
\bar{x}_1=\frac{\sqrt{\frac{C_1(d-1)}{d-2}} \left(2-\frac{(d-2)}{2 (d-1)} \frac{\lambda _2^2}{C_2}\right)}{2+\lambda _1} \quad \quad \quad \quad \bar{x}_2=\frac{\lambda_2}{2} \sqrt{\frac{d-2}{C_2(d-1)}} ,
\end{equation}
and
\begin{equation}
 \bar{y}_1= \pm \frac{\sqrt{\frac{C_1(d-1)}{d-2}} \left(2-\frac{(d-2)}{2 (d-1)} \frac{\lambda _2^2}{C_2}\right)}{2+\lambda _1} \sqrt{\frac{\lambda_1(2+\lambda_1)}{2\left(2-\frac{(d-2)}{2 (d-1)} \frac{\lambda _2^2}{C_2}\right)}\frac{d-2}{C_1(d-1)}-1}
 \quad \quad \quad \bar{y}_2=0.
\end{equation}
It exists for
\begin{equation}\label{eq:iniq}
 4 \frac{d-1}{d-2} - \frac{\lambda_1}{C_1}(2+\lambda_1) < \frac{\lambda_2^2}{C_2} < 4 \frac{d-1}{d-2},
\end{equation}
but the eigenvalues cannot be obtained analytically. Therefore, we resort to Eqs. \eqref{eq:condl} or \eqref{eq:stag} to conclude stability is achieved for
\begin{equation}\label{eq:2ms}
\lambda_2 < 2 C_2 \frac{d-1}{d-2} \frac{ 2-\frac{(d-2)}{2 (d-1)} \frac{\lambda _2^2}{C_2} }{2+\lambda _1}.
\end{equation}
In Fig.\ref{fig:l12}, we can see how this matches with the numerical calculation of the eigenvalues, for particular choices of $C_1$ and $C_2$.

Lastly, the critical point $A_{12}$ where both axions are rolling is
\begin{equation}
\bar{x}_i=\frac{2}{2+\lambda_1+\lambda_2} \sqrt{\frac{C_i(d-1)}{d-2}}
\end{equation}
and
\begin{equation}
  \bar{y}_i=\frac{2}{2+\lambda_1+\lambda_2} \sqrt{\frac{C_i(d-1)}{d-2}} \sqrt{\frac{\lambda_i(2+\lambda_1+\lambda_2)}{4}\frac{d-2}{C_i(d-1)}-1}.
\end{equation}
where $i=1,2$. It exists for
\begin{equation}\label{eq:axfp}
\frac{\lambda_i}{C_i} >4 \frac{d-1}{d-2} \frac{1}{2+\lambda_1+\lambda_2} \quad \quad \quad i=1,2.
\end{equation}
According to \eqref{eq:condl} -\eqref{eq:stag}, it is always stable whenever it exists. It is interesting to notice how \eqref{eq:axfp} for $i=2$ is the exact complement of \eqref{eq:2ms} (and for $i=1$ it is the complement of  \eqref{eq:2ms} with $\lambda_1 \leftrightarrow \lambda_2$), showing how the two attractors points can never coexist. 

To present an overview of the situation, we have plotted the regions of stability of the different fixed points in Fig.\ref{fig:l12}, using either a numerical calculation or analytical expression (whenever they exist) for the eigenvalues. These match nicely with the analytical expression \eqref{eq:2ms}-\eqref{eq:axfp} discussed above, derived with the machinery of Section \ref{ssc:gen}. It is also evident from the figure, and from the analytical formulas, that there is exactly one stable fixed point for every value of the parameters, \emph{i.e.} the coloured regions never overlap.

\subsection{Adding matter or radiation}\label{ssc:mrad}

For a more complete treatment, one should also consider the situation where, together with the chiral multiplet(s) $\Phi_{(i)}$, other sources of energy density (such as matter or radiation) are present. Indeed, that will be the case in any realistic cosmological scenario, for both early and late-time universe applications.

To do so, let us consider an unspecified fluid with an equation of state $P_{\gamma}=(\gamma-1) \rho_{\gamma}$ (with $0 \leq \gamma \leq 2$), obeying the equation of motion
\begin{equation}\label{eq:fl}
\dot{\rho_{\gamma}}+\gamma (d-1)H \rho_{\gamma} =0.
\end{equation}
Since the fluid is assumed to interact only gravitationally with the other components, the equations of motions \eqref{eq:eom} remain unchanged, while the Friedmann constraint equation becomes
\begin{equation}\label{eq:Hm}
(d-1)(d-2)H^2= \sum_{i=1}^N C_i \frac{\dot{s}^2+\dot{a}^2}{s^2} +2V(s_i)+2 \rho_{\gamma}.
\end{equation}
Differentiating with respect to time, and using \eqref{eq:fl}, one gets
\begin{equation}
\dot{H}= -(d-1)H^2 + \frac{2V(s_i)}{d-2}+ \frac{(2-\gamma)\rho_{\gamma}}{d-2}
\end{equation}
We can then use the notation in \ref{sc:gs}, and also define a new variable
\begin{equation}\label{eq:cstr1}
w = \frac{1}{H} \sqrt{\frac{2 \rho_{\gamma} }{(d-1)(d-2)}} \quad \quad \text{s.t.} \quad \quad \sum_{j=1}^N (x_j^2+y_j^2)+z^2+w^2=1.
\end{equation}
As usual, the full equations of motion can be recast as the system
\begin{equation}\label{eq:sysNm}
\left\{
\begin{aligned}
\frac{d x_i}{dM}&=- \sqrt{\frac{(d-1)(d-2)}{C_i}} \left[  y_i^2 - \frac{\lambda_i}{2}\left(1-\sum_{j=1}^N (x_j^2+y_j^2)-w^2\right)\right] \\
& \quad \quad \quad \quad \quad \quad \quad \quad \quad \quad \quad \quad -x_i(d-1)\left(1-\sum_{j=1}^N (x_j^2+y_j^2) -\frac{\gamma}{2} w^2 \right)\\
\frac{d y_i}{dM}&=  \sqrt{\frac{(d-1)(d-2)}{C_i}} x_i y_i-(d-1) \left(1-\sum_{j=1}^N (x_j^2+y_j^2) -\frac{\gamma}{2} w^2 \right)y_i\\
\frac{d w}{dM}&=(d-1)w \left[ \sum_{j=1}^N (x_j^2+y_j^2) +\frac{\gamma}{2} \left( w^2-1 \right) \right]
\end{aligned}\right.,
\end{equation}
For a single chiral multiplet ($N=1$), this is exactly the system studied in \cite{Cicoli:2020cfj,Cicoli:2020noz,Brinkmann:2022oxy}, modulo the notation change
\begin{equation}\label{eq:notc}
k_1=\frac{1}{\sqrt{C}} \quad \quad k_2=\frac{\lambda}{\sqrt{C}}  \quad \quad x_1=x \quad \quad x_2 = y \quad \quad y_1= z.
\end{equation}

An obvious class of fixed points is the one where $w=0$, in which case the system reduces to \eqref{eq:sysN}. In particular, the locations of such fixed points in the $x_i,y_i$ space are exactly as in \eqref{eq:fp1}-\eqref{eq:fp2}, as studied in the previous section. The stability analysis, however, requires a bit more care. If we denote the RHS terms in $f_x^i \equiv \frac{d x_i}{dN} $,
$f_y^i \equiv \frac{d y_i}{dN} $ and $f_w \equiv \frac{d w}{dN} $, one has
\begin{equation}
\frac{\partial {f_x^i }}{\partial w} \Bigg \rvert_{w=0}=\quad \frac{\partial {f_y^i }}{\partial w}\Bigg \rvert_{w=0} =0 \quad \quad \quad \quad \frac{\partial {f_w }}{\partial x_i} \Bigg \rvert_{w=0} = \quad \frac{\partial {f_w }}{\partial y_i} \Bigg \rvert_{w=0} =0.
\end{equation}
Therefore, the Jacobian matrix for the system \eqref{eq:sysNm} reduces to a block diagonal form, where the blocks are the Jacobian of the system \eqref{eq:sysN} and the single element $\frac{\partial {f_w }}{\partial w}$.
Fixed points with $\bar{w}=0$ are therefore stable if and only if the corresponding fixed point in \eqref{eq:fp1}-\eqref{eq:fp2} is stable, and
\begin{equation}
  \sum_{j=1}^N (\bar{x}_j^2+\bar{y}_j^2) < \frac{\gamma}{2}.
\end{equation}
For a fixed point of the form \eqref{eq:fp1}-\eqref{eq:fp2}, this reduces to
\begin{equation}\label{eq:irm}
\frac{\sum_{j=1}^K \lambda_j+\frac{d-2}{2(d-1)} \sum_{j=K+1}^N \frac{\lambda_j^2}{C_j} }{2 +\sum_{j=1}^K \lambda_j} < \frac{\gamma}{2},
\end{equation}
which correctly reproduces the result in \cite{Cicoli:2020cfj} in the specific case $d=4, N=1$ and $\gamma=1$. Notice that this condition could not be derived analytically (but only numerically) in Ref. \cite{Cicoli:2020cfj}, as the system was written in terms of $z$ (and not $w$) using the constraint \eqref{eq:cstr1}.

If $\bar{w} \neq 0,1$ the only other possibility is
\begin{equation}
  \sum_{j=1}^N (\bar{x}_j^2+\bar{y}_j^2) = \frac{\gamma}{2} \left( 1-\bar{w}^2 \right).
\end{equation}
Analogously to what we did in Subsection \ref{ssc:gen}, we can rearrange the $y_i$ so that
\begin{equation}
\bar{y}_i \neq 0 \quad \text{for} \quad i=1...K \quad \quad \text{and} \quad  \quad \bar{y}_i =0 \quad \text{for} \quad i=K+1...N.
\end{equation}
For now, let us also assume that $z\neq 0$. Then, it follows that
\begin{equation}\label{eq:fp1a}
\bar{x}_i= \sqrt{\frac{C_i(d-1)}{d-2}} \frac{2-\gamma}{2} \quad \quad \quad
\bar{y}_i=\frac{2-\gamma}{2} \sqrt{  \frac{\lambda_i}{2-\gamma}  \left( 1-\bar{w}^2 \right)-\frac{C_i(d-1)}{d-2}}
 \quad \text{for} \quad i \leq K, 
\end{equation}
and
\begin{equation}\label{eq:fp2a} 
\bar{x}_i=\frac{\lambda_i \left( 1-\bar{w}^2 \right)}{2}\sqrt{\frac{d-2}{C_i(d-1)}} \quad \quad \quad \quad \quad \quad  \bar{y}_i=0  \quad \quad \quad \quad \quad \quad\text{for} \quad i>K,
\end{equation}
which are highly reminiscent of \eqref{eq:fp1}-\eqref{eq:fp2}.
Combining these,
\begin{equation}
\frac{2-\gamma}{2} \sum_{j=1}^K \lambda_j+ \frac{d-2}{d-1}\left( 1-\bar{w}^2 \right) \sum_{j=K+1}^N \frac{\lambda_j^2}{2 C_i} = \gamma,
\end{equation}
which completely solves the systems. Of course, for a solution to be admissible various additional requirements must be satisifed, as the $\{\bar{x}_i,\bar{y}_i,\bar{w}\}$ must all be real and compliant with \eqref{eq:cstr1}. For simplicity, we will not treat all of these cases in detail, and only focus on a single modulus.

For a single modulus, one can calculate the eigenvalues of the Jacobian at every critical point analytically, as detailed in Appendix \ref{app:A3}. Here, we shall only state the results in the form of Table \ref{tab:cp}.
\begin{table}[ht]
\begin{tabular}{ | c | c | c | c | c| }
    \hline
    $x$ & $y$ & $w$ & Existence & Stability \\ \hline \hline
    $\pm 1$ & $0$ & $0$ & always & No \\ \hline
    $0$ & $0$ & $1$ & always & No \\ \hline
    $\frac{\lambda}{2} \sqrt{\frac{C(d-1)}{d-2}}$ & $0$ & $0$ & $\lambda^2 < 4\frac{C(d-1)}{d-2} $ & $\lambda^2 < 2 \gamma C \frac{d-1}{d-2} $ \\ \hline
    $\frac{\gamma}{\lambda} \sqrt{\frac{C(d-1)}{d-2}}$ & $0$ & $\sqrt{1-\frac{2 \gamma C}{\lambda^2} \frac{d-1}{d-2}}$ & $\lambda^2 < 2 \gamma C \frac{d-1}{d-2}$ & $\lambda > \frac{2 \gamma}{2-\gamma}$\\
    \hline
    $\frac{2}{2+\lambda} \sqrt{\frac{C(d-1)}{d-2}}$ &  $\frac{2}{2+\lambda}    \sqrt{\frac{\lambda(\lambda+2)}{4} - \frac{C(d-1)}{d-2}}$ & $0$ & $\lambda(\lambda+2) > 4 C \frac{d-1}{d-2} $ & $\lambda < \frac{2 \gamma}{2-\gamma}$ \\
    \hline
  \end{tabular}
  \caption{Critical points of the system \eqref{eq:sysNm}, for a single modulus ($N=1$). More details are presented in Appendix \ref{app:A3}.}
  \label{tab:cp}
\end{table}

These extend the results of \cite{Cicoli:2020cfj,Cicoli:2020noz,Brinkmann:2022oxy} to arbitrary values of $\gamma$ and $d$, and agree with them for $d=4$ and $\gamma=1$ using \eqref{eq:notc}. Moreover, the expressions we derived are fully analytical.

\subsection{Physics of the attractors}

Having established the conditions on the existence and the stability of the attractors \eqref{eq:fp1} and \eqref{eq:fp2} (in the absence of matter or radiation), let us now make a few remarks on their physical features. From the definitions \eqref{eq:xiyi}, it is immediate to see that the Hubble rate takes the form
\begin{equation}
H(t) = \frac{1}{(d-1)t} \frac{2+\sum_{j=1}^K \lambda_j}{\sum_{j=1}^K \lambda_j+ \frac{d-2}{2(d-1)} \sum_{j=K+1}^N \frac{\lambda_j^2}{C_j} },
\end{equation}
Therefore, these are power law cosmologies with a scale factor $\mathfrak{a}_s(t) = t^{\alpha} $ and
\begin{equation}
\alpha= \frac{2+\sum_{j=1}^K \lambda_j}{(d-1)\sum_{j=1}^K \lambda_j+ \frac{d-2}{2} \sum_{j=K+1}^N \frac{\lambda_j^2}{C_j} }.
\end{equation}
Comparing \eqref{eq:xiyi} with \eqref{eq:fp1} and \eqref{eq:fp2}, one also sees that $s_i(t) \sim t^{\beta_i} $ and $a_i(t) \sim t^{\gamma_i} $,
where the exponents are given by the following equations. If $i\leq K$,
\begin{equation}
\beta_i = \frac{2-\frac{d-2}{2(d-1)} \sum_{j=K+1}^N \frac{\lambda_j^2}{C_j}}{\sum_{j=1}^K \lambda_j+ \frac{d-2}{2(d-1)} \sum_{j=K+1}^N \frac{\lambda_j^2}{C_j} },
\end{equation}
 whereas for $i > K$
 \begin{equation}
 \beta_i=
 \frac{(d-2) \lambda_i}{2(d-1)} \frac{2+\sum_{j=1}^K \lambda_j}{\sum_{j=1}^K \lambda_j+ \frac{d-2}{2(d-1)} \sum_{j=K+1}^N \frac{\lambda_j^2}{C_j} }.
 \end{equation}
In both cases, the formulas correctly reduce to $\beta = 2/\lambda$ for $N=1$. Finally, for $i\leq K$ $\gamma_i=0$, and for $i>K$, $\gamma_i=\beta_i$. 

To achieve accelerated expansion, one would need a stable fixed point of the form in \eqref{eq:fp1}-\eqref{eq:fp2}, and therefore compatible with \eqref{eq:cond1} and \eqref{eq:condl},  which satisfies the additional condition
\begin{equation}\label{eq:accex}
2\sum_{j=1}^K \lambda_j +\sum_{j=K+1}^N \frac{\lambda_j^2}{C_j} < \frac{4}{d-2}.
\end{equation}
Moreover, \eqref{eq:accex} automatically implies \eqref{eq:irm} if $\gamma > \frac{2}{d-1}$ (as is the case for matter or radiation, for any $d>2$), so that accelerating solutions exist and are stable even when a fluid is present.
As in the analysis of \cite{Cicoli:2020cfj,Cicoli:2020noz,Brinkmann:2022oxy}, this shows that non-geodesic trajectories involving the axions can soften the requirements for accelerated expansion, at the price of strong curvature in field space. However, it is also clear that introducing more than one field makes the bound \eqref{eq:accex} harder to satisfy,\footnote{Intuitively, this resonates with the fact that all sources of kinetic energy contribute positively to the $\varepsilon$ parameter.} so that the single modulus examples presented in \cite{Cicoli:2020cfj,Cicoli:2020noz,Brinkmann:2022oxy} are still the most favourable to achieve accelerated expansion (notice that the $\lambda_i$ with $i\leq K$ cannot be  chosen negative, for consistency with \eqref{eq:fp1}). Indeed, we anticipate that in Section \ref{sec:app} we will not find any explicit examples satisfying \eqref{eq:accex} and where the axions play a role (meaning $K <N$) in both classes of string theory potentials motivating our analysis,\footnote{And also of the form \eqref{eq:potf}.} \emph{i.e.} at a large volume or for asymptotic limits in complex structure moduli space. This can be interpreted as a negative result concerning multi-field saxion-axion quintessence, in an admittedly limited context.

\section{Phenomenology}\label{sec:app}

In order to discuss possible phenomenological implications,
let us discuss more in the detail some explicit examples of the type motivated in Section \ref{sec:mot}.

\subsection{Large Volume models}\label{ssc:LVm}
The Large Volume Scenario \cite{Balasubramanian:2005zx,Conlon:2005ki} and variations thereof are some of the most prominent settings featuring a saxion-axion EFT of the type we have been studying. In this context, it was already discussed in \cite{Conlon:2022pnx,Apers:2022cyl,Cicoli:2023opf} how an early kination epoch could arise after inflation if the volume modulus were to start its evolution displaced from the minimum, where the potential is to a very good approximation a pure exponential. Since a universal volume axion is also present, however, it is natural to ask whether the attractor solutions discussed in this paper could play a role. 

In LVS, the complex structure moduli and the axiodilaton are stabilised at a parametrically higher scale than the K\"ahler moduli, so that the relevant ones are the latter. We can parametrize them as $T_i= \tau_i+ic_i$, where the real part $\tau_i$ denotes the 4-cycle volume, and the imaginary part $c_i$ the $C_3$ axion. In its simplest incarnation, the compactification space can be taken to be a Swiss-cheese Calabi-Yau manifold with only one very large 4-cycle $\tau_b\equiv \tau$ determining the extra dimensional volume, and a small blow-up cycle $\tau_s$. In that case, the LVS low energy theory is quite simple (See the discussion in \ref{ssc:LVS}): all the moduli except $T_b$ are integrated out, and one is left with the action
\begin{equation}
\mathcal{L}=\frac{3}{4 \tau^2} \partial_\mu \tau \partial^\mu \tau+\frac{3}{4 \tau^2} \partial_\mu c \partial^\mu c + \frac{V_0}{\tau^{ 9/2}}\left(A-(\ln \tau)^{3/2}\right),
\end{equation}
where we have used (somewhat unusually) non-canonically normalised fields in order to make contact with the notation of Section \ref{sec:mot}. From the above, it is easy to see that $C=\frac{3}{2}$ and $\lambda = \frac{9}{2}$. If the evolution starts around $\tau \sim \mathcal{O}(1)$, the logarithmic term can be neglected and we are in the situation of Section \ref{sec:3}. Without matter or radiation, the attractor is characterised by
\begin{equation}\label{eq:atLVS}
x = 6/13 \quad \quad  y=9/13 \quad \quad z=\frac{2}{\sqrt{13}}.
\end{equation}

A slightly more involved case is the fibred version of LVS considered in section \ref{ssc:LVS},  In other words, the effective theory for the moduli is, at leading order in a derivative expansion,\footnote{See Appendix B.1 of \cite{Conlon:2021cjk} for the diagonal and canonically normalised expression.}
\begin{equation}
\mathcal{L}= \frac{1}{4 \tau_1^2} \partial_{\mu} \tau_1 \partial^{\mu} \tau_1 +\frac{1}{2 \tau_2^2} \partial_{\mu} \tau_2 \partial^{\mu} \tau_2+\frac{1}{4 \tau_1^2} \partial_{\mu} c_1 \partial^{\mu} c_1 +\frac{1}{2 \tau_2^2} \partial_{\mu} c_2 \partial^{\mu} c_2 + V(\mathcal{V}\left(\tau_1,\tau_2)\right),
\end{equation}
where the functional form of $V(\mathcal{V})$ is the same as in LVS with a Swiss cheese manifold. For phenomenological purposes, the flat direction can eventually be lifted by string loop corrections, but this effect will be ignored here. In the notation of section \ref{sec:3}, this corresponds to a multi-modulus system with
\begin{equation}\label{eq:cl}
C_1= \frac{1}{2} \quad \quad C_2=1 \quad \quad \lambda_1 = 3 \quad \quad \lambda_2 = \frac{3}{2}.
\end{equation}
The situation is depicted on the left plot in Fig.\ref{fig:l12}, where the values of the parameters corresponding to \eqref{eq:cl} are indicated by the point P1. From \eqref{eq:cl}, and with the results of section \ref{sc:gs} , we can deduce that the only attractor is
\begin{equation}\label{eq:LVSf}
x_1 = \frac{2\sqrt{3}}{13} \quad \quad  y_1=\frac{\sqrt{66}}{13}\quad \quad x_2 = \frac{2\sqrt{6}}{13}\quad \quad y_2=\frac{15}{13} \quad \quad z^2=\frac{4}{13}.
\end{equation}
In both the fibred and non-fibred cases, the volume will grow slower than in pure kination, as
\begin{equation}
\mathcal{V} \sim t^{2/3}.
\end{equation}
Both of the solutions above fall short of being stable in presence of radiation, with $\gamma=\frac{4}{3}$ (for example, one would require $\lambda <4$ in the single field case, while $\lambda = \frac{9}{2}$ for LVS). In presence of radiation, the only stable attractors are the well-known tracker solutions of \cite{Wetterich:1987fm,Copeland:1997et,Ferreira:1997hj}, where the axions have zero velocity. Therefore, they could be relevant for the state of the early universe, but only if the axions' kinetic energy after inflation were much larger than the energy density contained in radiation, as discussed more in detail below. 

\subsubsection{Early Universe}

As emphasized in \cite{Conlon:2022pnx,Apers:2022cyl,Cicoli:2023opf,Apers:2024ffe}, the period between the end of inflation and the onset of Big Bang Nucleosynthesis (BBN) represents a unique opportunity for moduli and axions to have (co)-dominated the energy density of the universe, as it is insofar completely unconstrained by observations. From a (string) theoretic perspective, it is well motivated to consider alternative cosmological histories where the moduli play an important role. 
A concrete proposal was made in \cite{Conlon:2022pnx,Apers:2022cyl,Apers:2024ffe}, based on the evolution of the volume modulus in typical scenarios of moduli stabilisation (in particular LVS). In such models, the volume starts rolling along a steep potential and undergoes a phase of kination, is subsequently ``caught up" by radiation and converges to a tracker solution, until it starts oscillating around the minimum of the potential and finally decays, giving rise to reheating. Such initial conditions can be motivated from the fact that the string scale, which sets the magnitude of all other physical scales in the theory, depends on the volume as $m_s \sim \frac{M_P}{\sqrt{\mathcal{V}}}$, and can be significantly suppressed with respect to the Planck scale at the LVS minimum. If the evolution were to start close to the minimum, it would then seem hard to obtain any stringy realisation of inflation, which is conventionally assumed to take place at much higher scales $\Lambda_{\rm{inf}} \sim 10^{15} \, \rm{GeV}$ to achieve the observed spectrum of scalar perturbations.\footnote{See however \cite{German:2001tz} for an example of low-scale inflation.} Although we remain agnostic about the specific mechanism giving rise to inflation, a scenario where the inflaton is also identified with the volume (and therefore featuring an evolution of this kind) was studied in \cite{Conlon:2008cj}. 

The purpose of this section is to extend a similar picture in a more general context, and explore how it is modified whenever axions are taken into account. We have seen in Section \ref{ssc:mrad} that for certain values of the parameters (see in particular Eq. \eqref{eq:irm}), the inclusion of matter or radiation does not affect the stability of the critical points that exist in their absence. Such attractors can involve non-trivial evolution for the axions, and are reached even if the initial velocity of the axions is negligible. A crucial output of the above analysis is that, in LVS, the critical point with evolving axions is unstable in the presence of radiation (or matter), and would not appear particularly relevant for the evolution. Moreover, the stable attractor in that case is the one that would also exist if the axion were absent. There is, however, one exception. If at the end of inflation the kinetic energy stored in axions were much larger than that stored in radiation, the latter could be temporarily ignored, and the evolution could stay close to an axion critical point for a finite amount of time. Since the kinetic energy of any spectator axion and of radiation would be diluted as
\begin{equation} 
\frac{C}{2} \frac{\dot{a}^2}{s_{\rm{inf}}^2} \sim e^{-6 H_{\rm{inf}} t} \quad \quad \quad \rho_{\gamma}\sim e^{-4 H_{\rm{inf}} t},
\end{equation}
respectively during inflation,\footnote{With $s_{\rm{inf}}$ the value of the saxion during inflation.} such a scenario would require a sizeable axion velocity to be produced \emph{after} inflation. This is attainable in models where the axion(s) takes part in the inflationary dynamics, or at least has some non-trivial time evolution during inflation. See for example \cite{Bond:2006nc} for the K\"ahler axion in LVS, or \cite{Hebecker:2014kva} in the case of complex structure axions. As observed in \cite{Conlon:2022pnx}, a potentially dominant source of radiation for LVS is represented by perturbative decays of the volume into axions, so that the initial axion velocity $\frac{\dot{a}}{s_{\rm{inf}}}$  would have to satisfy \footnote{Notice that, although \eqref{eq:v0} requires $\dot{a}$ to be very large, the proper axion velocity (which also determines the kinetic energy) is $\frac{\dot{a}}{s_{\rm{inf}}}$, due to a non-canonical normalisation.}
\begin{equation}\label{eq:v0}
\frac{C}{2} \frac{\dot{a}^2(t_0)}{s_{\rm{inf}}^2}  \gg \alpha  \left( \frac{\lambda}{16 \pi ^3} \right)^3 \left( \frac{H_{\rm{inf}} }{2 \pi} \right)^2,
\end{equation}
with $\lambda$ the exponent in the LVS potential and $\alpha$ an $\mathcal{O}(1)$ branching factor. Notice that, if \eqref{eq:v0} were satisfied, during the early stages of the evolution the axion kinetic energy would also redshift slower than the radiation energy density
\begin{equation}
\rho_{\rm{ax}} \sim  \frac{\dot{a}^2}{s^2} \sim t^{-2/3} \quad \quad \quad \rho_{\gamma} \sim  \mathfrak{a}_s(t)^{-4} \sim t^{-4/3} ,
\end{equation}
as long as $\dot{a}^2 \ll \dot{s}^2$. Therefore, the system would first approach an (unstable) critical point of the form \eqref{eq:atLVS} (which we will also refer to as axion tracker), and only then would radiation begin to catch up. If, as in \cite{Conlon:2022pnx}, the volume started rolling immediately after inflation,\footnote{So that it could be considered kinating from a time $t_0$ satisfying $\frac{M_P^2}{3 t_0^2} \gtrsim V\left(\phi_0\right) \sim \Lambda_{\mathrm{inf}}^4$.} such a tracker solution would be reached approximately at
\begin{equation}
t_{\rm{tr},i} \simeq \frac{1}{3 H_{\rm{in}}} \left( \frac{C \dot{a}^2(t_0)}{3 s_{\rm{inf}}^2 H^2_{\rm{inf}}} \right)^{-3/4}.
\end{equation}
Once on the tracker, the relative growth of radiation would be slow, since the value of $\lambda$ for LVS is close to the boundary of the region where such a critical point is stable,
\begin{equation}
\frac{\rho_{\gamma}}{\rho_{\rm{ax}}} \sim t^{2-\frac{4}{3} \frac{2+\lambda}{\lambda}} \sim t^{2/27},
\end{equation}
and the system could stay close to \eqref{eq:atLVS} for a very long time.
As a quick estimate, 
\begin{equation}
t_{\rm{tr},f} \simeq t_{\rm{tr},i} \left( \frac{C \dot{a}^2(t_0)}{2 s_{\rm{inf}}^2 \rho_{\gamma,\rm{inf}}} \right)^{27/2} \left( \frac{t_{{\rm{tr,i}}}}{t_0} \right)^9,
\end{equation}
so that even with a moderate suppression of initial radiation such a tracker phase would be able to last for a significant portion of the universe' history before reheating. As an aside, a dynamics of this type could help to solve the overshoot problem \cite{Brustein:1992nk} as in \cite{Barreiro:1998aj,Conlon:2008cj,Conlon:2022pnx}, even in absence of initial radiation, but with the axion velocity playing the same role. 

Furthermore, axion tracker solutions have quite distinct features with respect to kination or radiation/matter trackers, for example in terms of the Hubble rate (and scale factor) or time dependence of the volume (and thus the string scale) in LVS. It would be extremely interesting to explore if epochs of this type could give rise to any observable traces, both in LVS or in more general contexts. In principle, a possible signature could be in terms of gravitational waves, either from an amplification of the inflationary spectrum (see \cite{Gouttenoire:2019kij,Co:2021lkc,Gouttenoire:2021wzu,Gouttenoire:2021jhk,Muia:2023wru} in the case of kination, although with a different origin) or from a fundamental string network \cite{Sarangi:2002yt,Copeland:2003bj}(See \cite{Conlon:2022pnx} for LVS).

\subsection{Other boundaries of moduli space}

A more general setting where potentials of the type \eqref{eq:potf} can be realised is provided by other boundaries of moduli space, as discussed (and motivated) in Section \ref{ssc:aht}. From Eqs \eqref{eq:kcs} and \eqref{eq:vcs}, the corresponding EFTs are characterised by a lagrangian of the form
\begin{equation}\label{eq:lb}
\mathcal{L} \supset \sum_{i=1}^N \frac{\Delta d_i}{2}\frac{\partial_{\mu} s^i \partial^{\mu}s^i+\partial_{\mu} a^i \partial^{\mu} a^i}{(s^i)^2} - V_0 \prod_{i=1}^N \frac{1}{s^{\Delta \ell_i}}.
\end{equation}
While lagrangians of this type are highly generic in string theory, a non-trivial output of the analysis of asymptotic potentials is that the precise numbers appearing in \eqref{eq:lb} are highly constrained, depending on the  type of singularity that is approached. While the complete set of possibilities is quite intricate, some basic consequences can already be drawn from the general features exposed in \ref{ssc:aht}. As an example, the kinetic term prefactors can only take the values
\begin{equation}
C_i = \frac{\Delta d_i}{2} \quad \quad \quad \quad \quad \Delta d_i \in \{ 0,1,2,3\},
\end{equation}
and if we only consider asymptotically vanishing potentials the corresponding exponents are constrained by
\begin{equation}
\lambda_i = \Delta \ell_i \quad \quad \quad \quad \quad \Delta \ell_i \in \{ 1,2,3,4\}.
\end{equation}
From \eqref{eq:accex}, it is then immediate to see how potentials of this type only admit asymptotic accelerated expansion with vanishing axion velocity, \emph{i.e.} for fixed points of the form \eqref{eq:fp2} with $K=0$.
This possibility was already discussed in \cite{Calderon-Infante:2022nxb}. Another generic consequence that can be drawn from the above is that, for a single modulus, any tracker solution with $w=0$ will be stable even in presence of radiation (with the potential exception of the marginal case $\lambda =4$).

\subsubsection{Late Universe}\label{ssc:lu}

Examples of complex structure moduli in F-theory were recently investigated in the context of the hunt for accelerated expansion in String Theory. In particular, Ref. \cite{Calderon-Infante:2022nxb} provided candidate\footnote{The K\"ahler moduli still need to be stabilised, otherwise they would provide a steep runaway direction which can spoil accelerated expansion.} examples of potentials at the boundary of moduli space to realise asymptotic accelerated expansion (See however \cite{Shiu:2023nph,Shiu:2023fhb,Hebecker:2023qke,VanRiet:2023cca} for arguments against asymptotic accelerated expansion). The point we would like to stress here is that, in the presence of axions, the true asymptotic solution (i.e. the exact attractor) might change, as we saw in the previous section. However, we should also remark that the solutions we have obtained only concern a very limited class of potentials, and in particular do not include cases with more than one term competing asymptotically, which are exactly those giving rise to accelerated expansion in \cite{Calderon-Infante:2022nxb}.

To see this, let us take one of the examples considered in \cite{Calderon-Infante:2022nxb}, which \emph{does not} realise accelerated expansion, but can be treated within our framework. It consists of two moduli, $T_s =s +ic$ and $T_u=u+ib$, with a lagrangian
\begin{equation}\label{eq:ex12}
\mathcal{L} = \frac{3}{4}\frac{(\partial s)^2+(\partial c)^2}{s^2}+\frac{1}{4}\frac{(\partial u)^2+(\partial b)^2}{u^2} -\frac{f_4^2}{u s}-\frac{(f_6\pm f_4 b)^2}{u^3 s},
\end{equation}
where $f_4,f_6$ are flux parameters that can be chosen arbitrarily. In \cite{Calderon-Infante:2022nxb}, the late-time behaviour is described by a gradient flow solution for the saxions, but we will now show that the presence of axions leads to a different late-time attractor.

Close to the boundary of moduli space, the second term in the potential is always subdominant\footnote{One can verify a posteriori that $u \gg b$.}, and therefore it can be neglected, so that the corresponding equations of motion are of the form discussed in Sections \ref{sc:gs} and \ref{ssc:tm}.
In our notation,
\begin{equation}\label{eq:fpac}
C_1 = \frac{1}{2} \quad \quad C_2 = \frac{3}{2} \quad \quad \lambda_1=1\quad \quad \lambda_2 = 1.
\end{equation}
Therefore, according to the analysis above, the attractor solution where both axions are constant is unstable, and the only stable attractor is characterised by
\begin{equation}\label{eq:solax}
x_1 = \frac{8 \sqrt{3}}{27}\quad \quad x_2 =\frac{1}{3} \quad \quad y_1=\frac{4}{27}\sqrt{\frac{3}{2}} \quad \quad y_2 =0.
\end{equation}
This can also be seen from the right plot in Fig.\ref{fig:l12}, where the point \eqref{eq:fpac} is marked as P2. To see the instability directly, we can also consider the system \eqref{eq:sys1} for the saxions in the approximation where we neglect the kinetic energy in the axions (and keeping only the dominant term in the potential):
\begin{equation}\label{eq:exeoms}
\left\{
\begin{aligned}
& \ddot{s}-\frac{\dot{s}^2}{s}+ 3 H \dot{s}-\frac{2 f_4^2}{u}=0\\
& \ddot{u}-\frac{\dot{u}^2}{u}+ 3 H \dot{u}-\frac{2 f_4^2}{u}=0\\
& 6H^2 =\frac{1}{2}\frac{\dot{s}^2}{s^2}+\frac{3}{2}\frac{\dot{u}^2}{u^2}+\frac{2 f_4^2}{us} \\
\end{aligned}\right.
\end{equation}
An exact solution, which is also the late time attractor without axions, is given by
\begin{equation}
s = \frac{4 f_4^2}{15 K} t^{3/2} \quad \quad u = K t^{1/2} \quad \quad H=\frac{3}{4t},
\end{equation}
with $K$ an integration constant. Such attractors are also known as scaling solutions, as all the terms in the equation of motion scale with the same power of $t$. For this reason, it is not consistent to perform a slow-roll approximation in \eqref{eq:exeoms}  and, for instance, neglect the acceleration term with respect to the friction one \cite{Shiu:2023nph,Shiu:2023fhb}.

From Eq. \eqref{eq:a0}, it can be readily seen that if we perturb the axions (assumed to be stabilised at their minimum, so with $\partial_a V \simeq 0$) with a very small initial velocity at a time $t_0$, they start growing as
\begin{equation}
\dot{c}(t) = \dot{c}(t_0) \left(\frac{t}{t_0}\right)^{3/4} \quad \quad \quad \dot{b}= \dot{b}(t_0) \left(\frac{t}{t_0}\right)^{-5/4}.
\end{equation}
While the second axion is indeed damped, the first one is unstable, as in \eqref{eq:solax}. Notice, in particular, that the axion $c$ is an exact flat direction of the potential \eqref{eq:ex12}, so its equation of motion is solved by \eqref{eq:a0} exactly. In turn, the backreaction of those axions will push the system out of this solution, and (for suitable initial conditions) converge to the attractor above.

As already mentioned, our analysis does not allow one to tackle the more interesting case of potentials with competing terms and determine the true asymptotic attractor(s) in such cases. However, the stability of the solutions for the saxions where the axions are stabilised at their minimum can still be investigated using Eq. \eqref{eq:a0}.
In particular, we can consider the examples in giving rise to asymptotic acceleration. Using the same notation as above, they are characterised by
\begin{equation}
\mathcal{L}_{\text{kin}} = \frac{1}{2}\frac{(\partial s)^2+(\partial c)^2}{s^2}+\frac{3}{2}\frac{(\partial u)^2+(\partial b)^2}{u^2} 
\end{equation}
and
\begin{equation}
V= f_2^2\frac{u}{s}+h_0^2 \frac{s}{u^3}+ \frac{(f_4\pm f_2 b)^2}{u s}+\frac{(f_6\pm f_4 b+f_2b^2-h_0 c)^2}{u^3 s},
\end{equation}
where $h_0,f_2,f_4$ and $f_6$ are again fluxes. These expression are valid within a growth sector $s \gg u \gg 1$, so that the last two terms are sub-dominant. As before, the trajectory of the system is determined by solving the full equations of motion
\begin{equation}
\left\{
\begin{aligned}
& \ddot{s}-\frac{\dot{s}^2}{s}+3 H \dot{s}-2 f_2^2 u+2\frac{h_0^2s^2}{u^3}=0\\
& \ddot{u}-\frac{\dot{u}^2}{u}+3 H \dot{u}+\frac{2 f_2^2}{3}\frac{u^2 }{s}-2 h_0^2\frac{s}{u^2}=0\\
& 6H^2 = \frac{1}{2}\frac{\dot{s}^2}{s^2}+\frac{3}{2}\frac{\dot{u}^2}{u^2}+2 f_2^2\frac{u}{s}+2 h_0^2 \frac{s}{u^3}. \\
\end{aligned}\right.
\end{equation}
It admits the particular solution
\begin{equation}
s =\frac{\abs{f_2^3 h_0}}{2700 \sqrt{5}} t^4 \quad \quad u =\frac{\abs{f_2 h_0}}{30 \sqrt{5}} t^2  \quad \quad H=\frac{7}{t},
\end{equation}
which is also a global attractor, as argued numerically \cite{Collinucci:2004iw} and recently proven in \cite{Shiu:2023fhb}. For the axions, if they are close to the minimum $\partial_{c,b} V \simeq 0$, and one can use \eqref{eq:a0} to conclude
\begin{equation}
\dot{c}(t) \simeq \dot{c}(t_0) \left(\frac{t}{t_0}\right)^{-13} \quad \quad \quad \dot{b} \simeq \dot{b}(t_0) \left(\frac{t}{t_0}\right)^{-17},
\end{equation}
so that the solution does not appear unstable. The other example of accelerated expansion in \cite{Calderon-Infante:2022nxb}, based on a singularity enhancement going beyond the weakly coupled limit in type IIB, also appears stable. In this case, the kinetic term and potential schematically read
\begin{equation}
\mathcal{L}_{\text{kin}} = \frac{(\partial s)^2+(\partial c)^2}{s^2}+\frac{(\partial u)^2+(\partial b)^2}{u^2}
\end{equation}
and
\begin{equation}
V \sim \frac{f^2}{u^2}+g^2\frac{u^2}{s^2}.
\end{equation}
The late-time attractor is given by
\begin{equation}
s =\frac{2 \abs{f g}}{13}\sqrt{\frac{2}{3}} t^2 \quad \quad u = \frac{2 \abs{f} }{\sqrt{39}} t \quad \quad H=\frac{5}{2t},
\end{equation}
and near the minimum of their potential the axions behave as
\begin{equation}
\dot{c}(t) \simeq \dot{c}(t_0) \left(\frac{t}{t_0}\right)^{-7/2} \quad \quad \quad \dot{b} \simeq \dot{b}(t_0) \left(\frac{t}{t_0}\right)^{-9/2}.
\end{equation}

\section{Outlook}\label{sec:con}

In this paper we have discussed the cosmological evolution of a system of $N$ saxions and axions coupled through their kinetic term, with the saxions rolling down a power-law potential\footnote{Corresponding to an exponential potential for canonically normalised scalars.} of the form \eqref{eq:potf}. Following \cite{Wetterich:1987fm,Copeland:1997et,Ferreira:1997hj,Barreiro:1999zs,Collinucci:2004iw,Shiu:2023nph,Shiu:2023fhb}, this was achieved by reformulating the equations of motions as an autonomous dynamical system, whose structure we analysed in Section \ref{sec:3}. In particular, we were able to analytically characterise all the critical points of the system and analyse their local stability for arbitrary $N$, bypassing a brute-force calculation of the eigenvalues of the Jacobian matrix. We also proved analytical results in the presence of an additional background fluid, such as matter or radiation.

We would like to emphasise that such saxion-axion systems are ubiquitous in low energy actions deriving from string theory, and the coupling arising from the kinetic term is a universal feature that characterizes them. For definiteness, we focused on two specific classes of examples originating from string theory: the Large Volume Scenario (and variations thereof), as well as asymptotic potentials at the boundary of moduli space. In Section \ref{sec:app}, we discussed situations where the coupled dynamics of saxions and axions could have played a significant role in the history of the universe. An intriguing possibility we discussed is that of an exotic cosmological epoch, such as kination or a tracker solution \cite{Conlon:2022pnx,Apers:2022cyl,Apers:2024ffe}, occurring between inflation and BBN. On a different note, we also commented on the role and stability of such solutions for accelerated expansion, building on the recent works \cite{Cicoli:2020cfj,Cicoli:2020noz,Brinkmann:2022oxy,Calderon-Infante:2022nxb}.

From a technical point of view, one might hope to extend our results to more complicated scenarios, such as the case where the saxion potential can be expressed as a sum of terms of the form \eqref{eq:potf}. In the absence of axions, such systems were thoroughly analysed in \cite{Shiu:2023nph,Shiu:2023fhb}, where the scaling solutions discussed in \cite{Collinucci:2004iw} were proven to be the global attractors under quite general assumptions. It would be interesting if similar statements could be generalised to include the axions, and perhaps even extended to the case where they have a non-zero potential. Any progress in either of these cases would allow a more complete characterisation of the cosmological behaviours attainable at the boundary of moduli space, allowing further investigation on, for example, accelerated expansion.

With a more phenomenological outlook, it was recently suggested that the QCD axion abundance from the misalignment mechanism could have been modified from a (large) initial axion velocity, a scenario known as kinetic misalignment \cite{Co:2019wyp,Co:2019jts,Chang:2019tvx} (See also \cite{Gouttenoire:2019kij,Co:2021lkc,Gouttenoire:2021wzu,Gouttenoire:2021jhk,Muia:2023wru} for some of the phenomenological consequences). While we have not even attempted to discuss how and if a proper QCD axion could arise in the example we mentioned (See \cite{Conlon:2006tq} for the QCD axion in LVS), it would be fascinating to explore whether coupled dynamics of saxions and axions similar to the one we have analysed could impact the abundance of a stringy QCD axion, and therefore the cosmological upper bound on its decay constant. Since with the standard misalignment mechanism the latter reads $f_a \lesssim 10^{12}$ GeV \cite{Abbott:1982af,GrillidiCortona:2015jxo}, one might wonder if it could be relaxed to approach typical values in string theory, $f_a \sim m_s \sim 10^{16}$ GeV \cite{Conlon:2006tq,Cicoli:2012sz}. \footnote{Although in the case of kinetic misalignment, a large initial velocity for the axion makes the bound more stringent.} As a more general, long term direction, it would be extremely interesting to investigate observational imprints (if any) of exotic, string-theory motivated cosmological epochs such as the ones studied in this work. This is a vast and largely unexplored territory, with potentially very high rewards. As mentioned in Section \ref{ssc:LVm}, some possibilities left to future analysis are the impact on cosmological perturbations \cite{Apers:2024ffe}, or modifications of primordial gravitational waves spectra \cite{Gouttenoire:2019kij,Co:2021lkc,Gouttenoire:2021wzu,Gouttenoire:2021jhk,Muia:2023wru}.

\subsection*{Acknoweldgements}

We are particularly grateful to Fien Apers, José Calderón-Infante, Joe Conlon, Ed Copeland, Thomas Grimm, Stefano Lanza, Flavio Tonioni and Damian van de Heisteeg for very valuable discussions, and comments on the manuscript. We would also like to thank the anonymous referee for many useful comments. The research of FR is supported by the Dutch Research Council (NWO) via a Start-Up grant and a Vici grant. 

\appendix
\section{Stability of the critical points}\label{app:A}
We present here some more details regarding the stability of the critical points, which is studied by linearising the system around them. In particular, the behaviour will be determined by the (sign of) the real part of the Hessian matrix's eigenvalues.
\subsection{One modulus}\label{app:a1}
The first two critical points correspond to pure saxion kination, namely
\begin{equation}
x= \pm 1 \quad \quad y=0 \quad \quad z=0.
\end{equation}
They exist for any values of $d$ and $\lambda$, with eigenvalues given by
\begin{equation}
\kappa_1^{\pm} = \pm \sqrt{\frac{(d-1)(d-2)}{C}} \quad \quad \lambda_2^{\pm} = \pm 2(d-1) \mp \lambda \sqrt{\frac{(d-1)(d-2)}{C}}  .
\end{equation}
respectively. Therefore, the first one ($+$) is a saddle point for $\lambda > 2 \sqrt{\frac{C(d-1)}{d-2}}$, and unstable otherwise. The second one ($-$) is always an unstable point. 

The third critical point is given by
\begin{equation}
x = \frac{\lambda}{2}\sqrt{\frac{d-2}{C(d-1)}} \quad \quad  y=0 \quad \quad z=1-\frac{(d-2) \lambda ^2}{4 (d-1) C},
\end{equation}
and corresponds to a well-known scaling solution for the saxion and a zero axion. It exists for $\lambda < 2 \sqrt{\frac{C(d-1)}{d-2}} $, with eigenvalues
\begin{equation}
\kappa_1 = \frac{\lambda(\lambda+2)(d-2)}{4 C}-(d-1) \quad \quad \kappa_2= \frac{\lambda^2(d-2)}{4 C}-(d-1).
\end{equation}
It is a stable node for $\lambda < \sqrt{1+\frac{4C(d-1)}{d-2}}-1$, and a saddle point otherwise. 

Finally, there is a last critical point which is specific of the saxion-axion system, and where both fields are evolving non-trivially. We refer to it as the kinating critical point, and it is characterised by
\begin{equation}\label{eq:fpa}
x= \frac{2}{2+\lambda} \sqrt{\frac{C(d-1)}{d-2}} \quad \quad \quad y= \frac{2}{2+\lambda}  \sqrt{\frac{d-1}{d-2}}  \sqrt{\frac{\lambda(\lambda+2)}{4} \frac{d-2}{d-1}-C} \quad \quad z=\frac{2}{\lambda+2},
\end{equation}
and exists for $\lambda > \sqrt{1+\frac{4C(d-1)}{d-2}}-1$. It has eigenvalues 
\begin{equation}
\kappa_{1,2} = \frac{-(d-1) \pm \sqrt{\frac{(d-1)}{C}\left[ (d-2) \lambda (\lambda+2)^2-C(d-1)(9+4\lambda) \right]}}{\lambda+2}.
\end{equation}
For
\begin{equation}\label{eq:spi}
\lambda^3(d-2)+4\lambda^2(d-2)+4\left(d-2-C(d-1)\right) \lambda -9C(d-1) >0,
\end{equation}
the term in the square root is purely imaginary, and the point is a stable spiral. 
If \eqref{eq:spi} is not satisfied, one can still have an attractor if the argument of the square root is between $0$ and $(d-1)^2$, that is
\begin{equation}\label{eq:stab}
8C(d-1) <\lambda^3(d-2)+4\lambda^2(d-2)+4\left(d-2-C(d-1)\right) \lambda < 9C(d-1).
\end{equation}
In that case, the critical point is a stable node. For $d>2$ (as we are assuming), this condition is satisfied if $\lambda >\sqrt{1+\frac{4C(d-1)}{d-2}}-1$. Therefore, this critical point is always an attractor whenever is exists. 
\subsection{Two moduli}\label{app:1b}

The critical locus corresponding to saxion kination is the unit circle in the $x_1-x_2$ plane,
\begin{equation}
x_1= t  \quad \quad  x_2 = \pm \sqrt{1-t^2} \quad \quad y_1=y_2=0, \quad \quad  t \in [0,1].
\end{equation}
There, the eigenvalues of the Hessian are 
\begin{multline}
\vec{\kappa}= \left\{0, \sqrt{\frac{(d-1)(d-2)}{C_1}} t, \pm\sqrt{\frac{(d-1)(d-2)}{C_2}} \sqrt{1-t^2}, \right. \\ \left.
 2(d-1) -  \lambda_1\sqrt{\frac{(d-1)(d-2)}{C_1}} t-  \mp \lambda_2\sqrt{\frac{(d-1)(d-2)}{C_2}} \sqrt{1-t^2} \right\},
\end{multline}
so that none of these points is ever an attractor. 
The critical point without axions is characterised by
\begin{equation}
x_1= \frac{\lambda_1}{2} \sqrt{\frac{d-2}{C_1(d-1)}} \quad \quad x_2= \frac{\lambda_2}{2} \sqrt{\frac{d-2}{C_2(d-1)}} \quad \quad y_1=y_2=0,
\end{equation}
and it exists only if
\begin{equation}
\frac{\lambda_1^2}{C_1}+\frac{\lambda_2^2}{C_2} \leq 4\frac{d-1}{d-2}.
\end{equation}
The eigenvalues of the Hessian in this case are
\begin{multline}
\vec{\kappa}=-\frac{d-2}{4} \times
 \left\{  4\frac{d-1}{d-2}-\frac{\lambda_1^2}{C_1}-\frac{\lambda_2^2}{C_2}, \,  4\frac{d-1}{d-2}-\frac{\lambda_1(2+\lambda_1)}{C_1}-\frac{\lambda_2^2}{C_2}, \right.\\ \left.
 4\frac{d-1}{d-2}-\frac{\lambda_1^2}{C_1}-\frac{\lambda_2^2}{C_2},
4\frac{d-1}{d-2}-\frac{\lambda_1^2}{C_1}-\frac{\lambda_2(2+\lambda_2)}{C_2} \right\} 
\end{multline}
Therefore, it is a stable node if
\begin{equation}
\frac{2\lambda_1}{C_1} < 4 \frac{d-1}{d-2} -  \frac{\lambda_1^2}{C_1}-\frac{\lambda_2^2}{C_2} \quad \quad \text{and}\quad \quad \frac{2\lambda_2}{C_2} < 4  \frac{d-1}{d-2} -  \frac{\lambda_1^2}{C_1} -\frac{\lambda_2^2}{C_2},
\end{equation}
and a saddle point otherwise. 
Unlike the one-modulus case, there are also hybrid critical points where fields in the two different multiplets exhibit a different behaviour. A pair is located at
\begin{equation}\label{eq:cm1}
x_1=\frac{\sqrt{\frac{C_1(d-1)}{d-2}} \left(2-\frac{(d-2)}{2 (d-1)} \frac{\lambda _2^2}{C_2}\right)}{2+\lambda _1} \quad \quad \quad \quad x_2=\frac{\lambda_2}{2} \sqrt{\frac{d-2}{C_2(d-1)}} ,
\end{equation}
and
\begin{equation}\label{eq:cm2}
 y_1= \pm \frac{\sqrt{\frac{C_1(d-1)}{d-2}} \left(2-\frac{(d-2)}{2 (d-1)} \frac{\lambda _2^2}{C_2}\right)}{2+\lambda _1} \sqrt{\frac{\lambda_1(2+\lambda_1)}{2\left(2-\frac{(d-2)}{2 (d-1)} \frac{\lambda _2^2}{C_2}\right)}\frac{d-2}{C_1(d-1)}-1}
 \quad \quad \quad y_2=0.
\end{equation}
An analogous pair can be obtained from the first through the permutation $x_1 \leftrightarrow x_2$, $y_1 \leftrightarrow y_2$ and $\lambda_1 \leftrightarrow \lambda_2$, and the conclusions will be identical. In particular, \eqref{eq:cm1}-\eqref{eq:cm2} can only exist if
\begin{equation}
 4 \frac{d-1}{d-2} - \frac{\lambda_1}{C_1}(2+\lambda_1) < \frac{\lambda_2^2}{C_2} < 4 \frac{d-1}{d-2}.
\end{equation}
Although we could not determine analytic expressions for all the eigenvalues, one of them is given by
\begin{equation}
\kappa = - \frac{4 C_2(d-1)-(d-2) \lambda_2(2+\lambda_1 +\lambda_2)}{2 C_2 (2+\lambda_1)},
\end{equation}
so that a necessary (but not sufficient) condition for stability is
\begin{equation}
\lambda_2 < 4 C_2 \frac{d-1}{d-2} \frac{1}{2+\lambda_1 +\lambda_2},
\end{equation}
which is indeed weaker than \eqref{eq:2ms}.

Finally, there is the critical point where all axions velocities are non-zero, given by
\begin{equation}
x_1=\frac{2}{2+\lambda_1+\lambda_2} \sqrt{\frac{C_1(d-1)}{d-2}}, \quad \quad \quad x_2=\frac{2}{2+\lambda_1+\lambda_2} \sqrt{\frac{C_2(d-1)}{d-2}},
\end{equation}
\begin{equation}
  y_1=\frac{2}{2+\lambda_1+\lambda_2} \sqrt{\frac{C_1(d-1)}{d-2}} \sqrt{\frac{\lambda_1(2+\lambda_1+\lambda_2)}{4}\frac{d-2}{C_1(d-1)}-1},
\end{equation}
and
\begin{equation}
y_2=\frac{2}{2+\lambda_1+\lambda_2} \sqrt{\frac{C_2(d-1)}{d-2}} \sqrt{\frac{\lambda_2(2+\lambda_1+\lambda_2)}{4}\frac{d-2}{C_2(d-1)}-1}.
\end{equation}
It exists for
\begin{equation}
\frac{\lambda_1}{C_1}(2+\lambda_1+\lambda_2) > 4 \frac{d-1}{d-2} \quad \quad \text{and} \quad \quad \frac{\lambda_2}{C_2}(2+\lambda_1+\lambda_2) > 4 \frac{d-1}{d-2}.
\end{equation}
but we were not able to derive any explicit expressions for the eigenvalues.

\subsection{One modulus plus a background fluid}\label{app:A3}

The first two critical points correspond to pure saxion kination, namely
\begin{equation}
x= \pm 1 \quad \quad y=0 \quad \quad z=0 \quad \quad w=0.
\end{equation}
They exist for any values of $d$ and $\lambda$, with eigenvalues given by
\begin{equation}
 \vec{\kappa}_{\pm}= \left\{ \pm \sqrt{\frac{(d-1)(d-2)}{C}}, \pm 2(d-1) \mp \lambda \sqrt{\frac{(d-1)(d-2)}{C}} , \kappa_3=\frac{1}{2}(d-1)(2-\gamma) \right\}
\end{equation}
Therefore, the first one ($+$) is a saddle point for $\lambda > 2 \sqrt{\frac{C(d-1)}{d-2}}$, and unstable otherwise. The second one ($-$) is always a saddle. Another saddle point is the one dominated by the fluid, \emph{i.e.}
\begin{equation}
x= 0 \quad \quad y=0 \quad \quad z=0 \quad \quad w=1,
\end{equation}
with eigenvalues
\begin{equation}
\vec{\kappa}= \left\{ \gamma (d-1), -\frac{d-1}{2}, -\frac{(d-1)(2-\gamma)}{2}\right\}.
\end{equation}
The critical point corresponding to a scaling solution for the saxion is
\begin{equation}
x= \frac{\lambda}{2} \sqrt{\frac{d-2}{C(d-1)}} \quad \quad y=0 \quad \quad z=\sqrt{1-\frac{\lambda^2}{4C} \frac{d-2}{d-1}} \quad \quad w= 0
\end{equation}
with eigenvalues given by
\begin{equation}
\vec{\kappa}= \left\{\frac{(d-2) \lambda^2-2 \gamma  C (d-1)}{4 C},\frac{(d-2) \lambda  (\lambda +2)}{4 C}-d+1,\frac{(d-2) \lambda ^2}{4 C}-(d-1) \right\} 
\end{equation}
Therefore, it exists for $\lambda<2\sqrt{\frac{C(d-1)}{d-2}}$, and is stable for
\begin{equation}\label{eq:stabt}
\lambda^2 < 2 \gamma C \frac{d-1}{d-2}  \quad \quad \text{and}\quad \quad \lambda (\lambda+2) < 4C \frac{d-1}{d-2},
\end{equation}
the latter of which is redundant. The last critical point with a vanishing axion velocity is 
\begin{equation}
x= \frac{\gamma}{\lambda} \sqrt{\frac{C(d-1)}{d-2}} \quad \quad y=0 \quad \quad z= \sqrt{\frac{\gamma(2-\gamma)}{\lambda^2} \frac{C(d-1)}{d-2}} \quad \quad  w=\sqrt{1-\frac{2 \gamma C}{\lambda^2} \frac{d-1}{d-2}}
\end{equation}
which exists for
\begin{equation}\label{eq:statr}
\lambda^2 > 2 \gamma C \frac{d-1}{d-2}.
\end{equation}
Notice how this is exactly the complement of the second stability condition in \eqref{eq:stabt} above. The eigenvalues are
\begin{equation}
\vec{\kappa}= \left\{ (d-1)(\gamma (2+\lambda)-2 \lambda),-\frac{d-1}{4 \lambda} \left( (2-\gamma) \pm \sqrt{\left(2-\gamma\right)\left( 16 \frac{C \gamma^2}{\lambda^2}\frac{d-1}{d-2}-(9 \gamma-2)\right)} \right) \right\},
\end{equation}
so that stability is guaranteed if \eqref{eq:statr} and
\begin{equation}\label{eq:comps}
\lambda > \frac{2 \gamma}{2-\gamma}
\end{equation}
are verified. Finally, one has the non-geodesic fixed point, where both the saxion and the axion are evolving:
\begin{equation}
x= \frac{2}{2+\lambda} \sqrt{\frac{C(d-1)}{d-2}} \quad \quad \quad y= \frac{2}{2+\lambda}  \sqrt{\frac{C(d-1)}{d-2}}  \sqrt{\frac{\lambda(\lambda+2)}{4} \frac{d-2}{C(d-1)}-1}
\end{equation}
and
\begin{equation}
z= \sqrt{\frac{2}{\lambda+2}} \quad \quad w=0.
\end{equation}
It exists for
\begin{equation}\label{eq:stabla}
 4 C \frac{d-1}{d-2} <  \lambda(\lambda+2),
\end{equation}
with eigenvalues
\begin{equation}
\vec{\kappa}= \left\{ -\frac{(d-1) (\gamma  (\lambda +2)-2 \lambda )}{2 (\lambda +2)},
-(d-1)\frac{1 \pm \sqrt{4 \lambda +9-\frac{(d-2) \lambda  (\lambda +2)^2}{C (d-1)}}}{\lambda +2} \right\}.
\end{equation}
It is stable if \eqref{eq:stabla} and the complement of \eqref{eq:comps} hold,
\begin{equation}
\lambda < \frac{2 \gamma}{2-\gamma}.
\end{equation}
We notice how these results correctly map to those in \cite{Cicoli:2020cfj,Cicoli:2020noz,Brinkmann:2022oxy} with the dictionary \eqref{eq:notc}, and extend them for arbitrary values of $\gamma$ and $d$.
\bibliography{biblist}

\providecommand{\href}[2]{#2}\begingroup\raggedright\begin{thebibliography}{100}

\bibitem{Susskind:2003kw}
L.~Susskind, \emph{{The Anthropic landscape of string theory}},
  \href{https://arxiv.org/abs/hep-th/0302219}{{\ttfamily hep-th/0302219}}.

\bibitem{Vafa:2005ui}
C.~Vafa, \emph{{The String landscape and the swampland}},
  \href{https://arxiv.org/abs/hep-th/0509212}{{\ttfamily hep-th/0509212}}.

\bibitem{Ooguri:2006in}
H.~Ooguri and C.~Vafa, \emph{{On the Geometry of the String Landscape and the
  Swampland}},
  \href{https://doi.org/10.1016/j.nuclphysb.2006.10.033}{\emph{Nucl. Phys. B}
  {\bfseries 766} (2007) 21}
  [\href{https://arxiv.org/abs/hep-th/0605264}{{\ttfamily hep-th/0605264}}].

\bibitem{Palti:2019pca}
E.~Palti, \emph{{The Swampland: Introduction and Review}},
  \href{https://doi.org/10.1002/prop.201900037}{\emph{Fortsch. Phys.}
  {\bfseries 67} (2019) 1900037}
  [\href{https://arxiv.org/abs/1903.06239}{{\ttfamily 1903.06239}}].

\bibitem{Danielsson:2018ztv}
U.~H. Danielsson and T.~Van~Riet, \emph{{What if string theory has no de Sitter
  vacua?}}, \href{https://doi.org/10.1142/S0218271818300070}{\emph{Int. J. Mod.
  Phys. D} {\bfseries 27} (2018) 1830007}
  [\href{https://arxiv.org/abs/1804.01120}{{\ttfamily 1804.01120}}].

\bibitem{Obied:2018sgi}
G.~Obied, H.~Ooguri, L.~Spodyneiko and C.~Vafa, \emph{{De Sitter Space and the
  Swampland}},  \href{https://arxiv.org/abs/1806.08362}{{\ttfamily
  1806.08362}}.

\bibitem{Dine:1985he}
M.~Dine and N.~Seiberg, \emph{{Is the Superstring Weakly Coupled?}},
  \href{https://doi.org/10.1016/0370-2693(85)90927-X}{\emph{Phys. Lett. B}
  {\bfseries 162} (1985) 299}.

\bibitem{VanRiet:2023pnx}
T.~Van~Riet and G.~Zoccarato, \emph{{Beginners lectures on flux
  compactifications and related Swampland topics}},
  \href{https://arxiv.org/abs/2305.01722}{{\ttfamily 2305.01722}}.

\bibitem{Grana:2005jc}
M.~Grana, \emph{{Flux compactifications in string theory: A Comprehensive
  review}}, \href{https://doi.org/10.1016/j.physrep.2005.10.008}{\emph{Phys.
  Rept.} {\bfseries 423} (2006) 91}
  [\href{https://arxiv.org/abs/hep-th/0509003}{{\ttfamily hep-th/0509003}}].

\bibitem{Douglas:2006es}
M.~R. Douglas and S.~Kachru, \emph{{Flux compactification}},
  \href{https://doi.org/10.1103/RevModPhys.79.733}{\emph{Rev. Mod. Phys.}
  {\bfseries 79} (2007) 733}
  [\href{https://arxiv.org/abs/hep-th/0610102}{{\ttfamily hep-th/0610102}}].

\bibitem{Cicoli:2023opf}
M.~Cicoli, J.~P. Conlon, A.~Maharana, S.~Parameswaran, F.~Quevedo and
  I.~Zavala, \emph{{String Cosmology: from the Early Universe to Today}},
  \href{https://arxiv.org/abs/2303.04819}{{\ttfamily 2303.04819}}.

\bibitem{Coughlan:1983ci}
G.~D. Coughlan, W.~Fischler, E.~W. Kolb, S.~Raby and G.~G. Ross,
  \emph{{Cosmological Problems for the Polonyi Potential}},
  \href{https://doi.org/10.1016/0370-2693(83)91091-2}{\emph{Phys. Lett. B}
  {\bfseries 131} (1983) 59}.

\bibitem{Banks:1993en}
T.~Banks, D.~B. Kaplan and A.~E. Nelson, \emph{{Cosmological implications of
  dynamical supersymmetry breaking}},
  \href{https://doi.org/10.1103/PhysRevD.49.779}{\emph{Phys. Rev. D} {\bfseries
  49} (1994) 779} [\href{https://arxiv.org/abs/hep-ph/9308292}{{\ttfamily
  hep-ph/9308292}}].

\bibitem{deCarlos:1993wie}
B.~de~Carlos, J.~A. Casas, F.~Quevedo and E.~Roulet, \emph{{Model independent
  properties and cosmological implications of the dilaton and moduli sectors of
  4-d strings}},
  \href{https://doi.org/10.1016/0370-2693(93)91538-X}{\emph{Phys. Lett. B}
  {\bfseries 318} (1993) 447}
  [\href{https://arxiv.org/abs/hep-ph/9308325}{{\ttfamily hep-ph/9308325}}].

\bibitem{Svrcek:2006yi}
P.~Svrcek and E.~Witten, \emph{{Axions In String Theory}},
  \href{https://doi.org/10.1088/1126-6708/2006/06/051}{\emph{JHEP} {\bfseries
  06} (2006) 051} [\href{https://arxiv.org/abs/hep-th/0605206}{{\ttfamily
  hep-th/0605206}}].

\bibitem{Conlon:2006tq}
J.~P. Conlon, \emph{{The QCD axion and moduli stabilisation}},
  \href{https://doi.org/10.1088/1126-6708/2006/05/078}{\emph{JHEP} {\bfseries
  05} (2006) 078} [\href{https://arxiv.org/abs/hep-th/0602233}{{\ttfamily
  hep-th/0602233}}].

\bibitem{Arvanitaki:2009fg}
A.~Arvanitaki, S.~Dimopoulos, S.~Dubovsky, N.~Kaloper and J.~March-Russell,
  \emph{{String Axiverse}},
  \href{https://doi.org/10.1103/PhysRevD.81.123530}{\emph{Phys. Rev. D}
  {\bfseries 81} (2010) 123530}
  [\href{https://arxiv.org/abs/0905.4720}{{\ttfamily 0905.4720}}].

\bibitem{Jaeckel:2010ni}
J.~Jaeckel and A.~Ringwald, \emph{{The Low-Energy Frontier of Particle
  Physics}},
  \href{https://doi.org/10.1146/annurev.nucl.012809.104433}{\emph{Ann. Rev.
  Nucl. Part. Sci.} {\bfseries 60} (2010) 405}
  [\href{https://arxiv.org/abs/1002.0329}{{\ttfamily 1002.0329}}].

\bibitem{Cicoli:2012sz}
M.~Cicoli, M.~Goodsell and A.~Ringwald, \emph{{The type IIB string axiverse and
  its low-energy phenomenology}},
  \href{https://doi.org/10.1007/JHEP10(2012)146}{\emph{JHEP} {\bfseries 10}
  (2012) 146} [\href{https://arxiv.org/abs/1206.0819}{{\ttfamily 1206.0819}}].

\bibitem{Peccei:1977hh}
R.~D. Peccei and H.~R. Quinn, \emph{{CP Conservation in the Presence of
  Instantons}}, \href{https://doi.org/10.1103/PhysRevLett.38.1440}{\emph{Phys.
  Rev. Lett.} {\bfseries 38} (1977) 1440}.

\bibitem{Wilczek:1977pj}
F.~Wilczek, \emph{{Problem of Strong $P$ and $T$ Invariance in the Presence of
  Instantons}}, \href{https://doi.org/10.1103/PhysRevLett.40.279}{\emph{Phys.
  Rev. Lett.} {\bfseries 40} (1978) 279}.

\bibitem{Preskill:1982cy}
J.~Preskill, M.~B. Wise and F.~Wilczek, \emph{{Cosmology of the Invisible
  Axion}}, \href{https://doi.org/10.1016/0370-2693(83)90637-8}{\emph{Phys.
  Lett. B} {\bfseries 120} (1983) 127}.

\bibitem{Abbott:1982af}
L.~F. Abbott and P.~Sikivie, \emph{{A Cosmological Bound on the Invisible
  Axion}}, \href{https://doi.org/10.1016/0370-2693(83)90638-X}{\emph{Phys.
  Lett. B} {\bfseries 120} (1983) 133}.

\bibitem{Dine:1982ah}
M.~Dine and W.~Fischler, \emph{{The Not So Harmless Axion}},
  \href{https://doi.org/10.1016/0370-2693(83)90639-1}{\emph{Phys. Lett. B}
  {\bfseries 120} (1983) 137}.

\bibitem{Hui:2016ltb}
L.~Hui, J.~P. Ostriker, S.~Tremaine and E.~Witten, \emph{{Ultralight scalars as
  cosmological dark matter}},
  \href{https://doi.org/10.1103/PhysRevD.95.043541}{\emph{Phys. Rev. D}
  {\bfseries 95} (2017) 043541}
  [\href{https://arxiv.org/abs/1610.08297}{{\ttfamily 1610.08297}}].

\bibitem{Cicoli:2021gss}
M.~Cicoli, V.~Guidetti, N.~Righi and A.~Westphal, \emph{{Fuzzy Dark Matter
  candidates from string theory}},
  \href{https://doi.org/10.1007/JHEP05(2022)107}{\emph{JHEP} {\bfseries 05}
  (2022) 107} [\href{https://arxiv.org/abs/2110.02964}{{\ttfamily
  2110.02964}}].

\bibitem{Silverstein:2008sg}
E.~Silverstein and A.~Westphal, \emph{{Monodromy in the CMB: Gravity Waves and
  String Inflation}},
  \href{https://doi.org/10.1103/PhysRevD.78.106003}{\emph{Phys. Rev. D}
  {\bfseries 78} (2008) 106003}
  [\href{https://arxiv.org/abs/0803.3085}{{\ttfamily 0803.3085}}].

\bibitem{McAllister:2008hb}
L.~McAllister, E.~Silverstein and A.~Westphal, \emph{{Gravity Waves and Linear
  Inflation from Axion Monodromy}},
  \href{https://doi.org/10.1103/PhysRevD.82.046003}{\emph{Phys. Rev. D}
  {\bfseries 82} (2010) 046003}
  [\href{https://arxiv.org/abs/0808.0706}{{\ttfamily 0808.0706}}].

\bibitem{Kachru:2003aw}
S.~Kachru, R.~Kallosh, A.~D. Linde and S.~P. Trivedi, \emph{{De Sitter vacua in
  string theory}},
  \href{https://doi.org/10.1103/PhysRevD.68.046005}{\emph{Phys. Rev. D}
  {\bfseries 68} (2003) 046005}
  [\href{https://arxiv.org/abs/hep-th/0301240}{{\ttfamily hep-th/0301240}}].

\bibitem{Balasubramanian:2005zx}
V.~Balasubramanian, P.~Berglund, J.~P. Conlon and F.~Quevedo,
  \emph{{Systematics of moduli stabilisation in Calabi-Yau flux
  compactifications}},
  \href{https://doi.org/10.1088/1126-6708/2005/03/007}{\emph{JHEP} {\bfseries
  03} (2005) 007} [\href{https://arxiv.org/abs/hep-th/0502058}{{\ttfamily
  hep-th/0502058}}].

\bibitem{Conlon:2005ki}
J.~P. Conlon, F.~Quevedo and K.~Suruliz, \emph{{Large-volume flux
  compactifications: Moduli spectrum and D3/D7 soft supersymmetry breaking}},
  \href{https://doi.org/10.1088/1126-6708/2005/08/007}{\emph{JHEP} {\bfseries
  08} (2005) 007} [\href{https://arxiv.org/abs/hep-th/0505076}{{\ttfamily
  hep-th/0505076}}].

\bibitem{Cicoli:2018kdo}
M.~Cicoli, S.~De~Alwis, A.~Maharana, F.~Muia and F.~Quevedo, \emph{{De Sitter
  vs Quintessence in String Theory}},
  \href{https://doi.org/10.1002/prop.201800079}{\emph{Fortsch. Phys.}
  {\bfseries 67} (2019) 1800079}
  [\href{https://arxiv.org/abs/1808.08967}{{\ttfamily 1808.08967}}].

\bibitem{Hebecker:2019csg}
A.~Hebecker, T.~Skrzypek and M.~Wittner, \emph{{The $F$-term Problem and other
  Challenges of Stringy Quintessence}},
  \href{https://doi.org/10.1007/JHEP11(2019)134}{\emph{JHEP} {\bfseries 11}
  (2019) 134} [\href{https://arxiv.org/abs/1909.08625}{{\ttfamily
  1909.08625}}].

\bibitem{ValeixoBento:2020ujr}
B.~Valeixo~Bento, D.~Chakraborty, S.~L. Parameswaran and I.~Zavala, \emph{{Dark
  Energy in String Theory}},
  \href{https://doi.org/10.22323/1.376.0123}{\emph{PoS} {\bfseries CORFU2019}
  (2020) 123} [\href{https://arxiv.org/abs/2005.10168}{{\ttfamily
  2005.10168}}].

\bibitem{Cicoli:2020cfj}
M.~Cicoli, G.~Dibitetto and F.~G. Pedro, \emph{{New accelerating solutions in
  late-time cosmology}},
  \href{https://doi.org/10.1103/PhysRevD.101.103524}{\emph{Phys. Rev. D}
  {\bfseries 101} (2020) 103524}
  [\href{https://arxiv.org/abs/2002.02695}{{\ttfamily 2002.02695}}].

\bibitem{Cicoli:2020noz}
M.~Cicoli, G.~Dibitetto and F.~G. Pedro, \emph{{Out of the Swampland with
  Multifield Quintessence?}},
  \href{https://doi.org/10.1007/JHEP10(2020)035}{\emph{JHEP} {\bfseries 10}
  (2020) 035} [\href{https://arxiv.org/abs/2007.11011}{{\ttfamily
  2007.11011}}].

\bibitem{Brinkmann:2022oxy}
M.~Brinkmann, M.~Cicoli, G.~Dibitetto and F.~G. Pedro, \emph{{Stringy
  multifield quintessence and the Swampland}},
  \href{https://doi.org/10.1007/JHEP11(2022)044}{\emph{JHEP} {\bfseries 11}
  (2022) 044} [\href{https://arxiv.org/abs/2206.10649}{{\ttfamily
  2206.10649}}].

\bibitem{Cicoli:2021fsd}
M.~Cicoli, F.~Cunillera, A.~Padilla and F.~G. Pedro, \emph{{Quintessence and
  the Swampland: The Parametrically Controlled Regime of Moduli Space}},
  \href{https://doi.org/10.1002/prop.202200009}{\emph{Fortsch. Phys.}
  {\bfseries 70} (2022) 2200009}
  [\href{https://arxiv.org/abs/2112.10779}{{\ttfamily 2112.10779}}].

\bibitem{Cicoli:2021skd}
M.~Cicoli, F.~Cunillera, A.~Padilla and F.~G. Pedro, \emph{{Quintessence and
  the Swampland: The Numerically Controlled Regime of Moduli Space}},
  \href{https://doi.org/10.1002/prop.202200008}{\emph{Fortsch. Phys.}
  {\bfseries 70} (2022) 2200008}
  [\href{https://arxiv.org/abs/2112.10783}{{\ttfamily 2112.10783}}].

\bibitem{Christodoulidis:2019jsx}
P.~Christodoulidis, D.~Roest and E.~I. Sfakianakis, \emph{{Scaling attractors
  in multi-field inflation}},
  \href{https://doi.org/10.1088/1475-7516/2019/12/059}{\emph{JCAP} {\bfseries
  12} (2019) 059} [\href{https://arxiv.org/abs/1903.06116}{{\ttfamily
  1903.06116}}].

\bibitem{Christodoulidis:2021vye}
P.~Christodoulidis and A.~Paliathanasis, \emph{{$\mathcal{N}$-field cosmology
  in hyperbolic field space: stability and general solutions}},
  \href{https://doi.org/10.1088/1475-7516/2021/05/038}{\emph{JCAP} {\bfseries
  05} (2021) 038} [\href{https://arxiv.org/abs/2101.09582}{{\ttfamily
  2101.09582}}].

\bibitem{Christodoulidis:2022vww}
P.~Christodoulidis and R.~Rosati, \emph{{(Slow-)twisting inflationary
  attractors}},
  \href{https://doi.org/10.1088/1475-7516/2023/09/034}{\emph{JCAP} {\bfseries
  09} (2023) 034} [\href{https://arxiv.org/abs/2210.14900}{{\ttfamily
  2210.14900}}].

\bibitem{Rudelius:2022gbz}
T.~Rudelius, \emph{{Asymptotic scalar field cosmology in string theory}},
  \href{https://doi.org/10.1007/JHEP10(2022)018}{\emph{JHEP} {\bfseries 10}
  (2022) 018} [\href{https://arxiv.org/abs/2208.08989}{{\ttfamily
  2208.08989}}].

\bibitem{Calderon-Infante:2022nxb}
J.~Calder\'on-Infante, I.~Ruiz and I.~Valenzuela, \emph{{Asymptotic accelerated
  expansion in string theory and the Swampland}},
  \href{https://doi.org/10.1007/JHEP06(2023)129}{\emph{JHEP} {\bfseries 06}
  (2023) 129} [\href{https://arxiv.org/abs/2209.11821}{{\ttfamily
  2209.11821}}].

\bibitem{Cremonini:2023suw}
S.~Cremonini, E.~Gonzalo, M.~Rajaguru, Y.~Tang and T.~Wrase, \emph{{On
  asymptotic dark energy in string theory}},
  \href{https://doi.org/10.1007/JHEP09(2023)075}{\emph{JHEP} {\bfseries 09}
  (2023) 075} [\href{https://arxiv.org/abs/2306.15714}{{\ttfamily
  2306.15714}}].

\bibitem{Freigang:2023ogu}
J.~Freigang, D.~Lust, G.-E. Nian and M.~Scalisi, \emph{{Cosmic acceleration and
  turns in the Swampland}},
  \href{https://doi.org/10.1088/1475-7516/2023/11/080}{\emph{JCAP} {\bfseries
  11} (2023) 080} [\href{https://arxiv.org/abs/2306.17217}{{\ttfamily
  2306.17217}}].

\bibitem{Andriot:2023wvg}
D.~Andriot, D.~Tsimpis and T.~Wrase, \emph{{Accelerated expansion of an open
  universe and string theory realizations}},
  \href{https://doi.org/10.1103/PhysRevD.108.123515}{\emph{Phys. Rev. D}
  {\bfseries 108} (2023) 123515}
  [\href{https://arxiv.org/abs/2309.03938}{{\ttfamily 2309.03938}}].

\bibitem{Antoniadis:1988vi}
I.~Antoniadis, C.~Bachas, J.~R. Ellis and D.~V. Nanopoulos, \emph{{An Expanding
  Universe in String Theory}},
  \href{https://doi.org/10.1016/0550-3213(89)90095-3}{\emph{Nucl. Phys. B}
  {\bfseries 328} (1989) 117}.

\bibitem{Dvali:1998pa}
G.~R. Dvali and S.~H.~H. Tye, \emph{{Brane inflation}},
  \href{https://doi.org/10.1016/S0370-2693(99)00132-X}{\emph{Phys. Lett. B}
  {\bfseries 450} (1999) 72}
  [\href{https://arxiv.org/abs/hep-ph/9812483}{{\ttfamily hep-ph/9812483}}].

\bibitem{Choi:1999xn}
K.~Choi, \emph{{String or M theory axion as a quintessence}},
  \href{https://doi.org/10.1103/PhysRevD.62.043509}{\emph{Phys. Rev. D}
  {\bfseries 62} (2000) 043509}
  [\href{https://arxiv.org/abs/hep-ph/9902292}{{\ttfamily hep-ph/9902292}}].

\bibitem{Kachru:2003sx}
S.~Kachru, R.~Kallosh, A.~D. Linde, J.~M. Maldacena, L.~P. McAllister and S.~P.
  Trivedi, \emph{{Towards inflation in string theory}},
  \href{https://doi.org/10.1088/1475-7516/2003/10/013}{\emph{JCAP} {\bfseries
  10} (2003) 013} [\href{https://arxiv.org/abs/hep-th/0308055}{{\ttfamily
  hep-th/0308055}}].

\bibitem{Conlon:2005jm}
J.~P. Conlon and F.~Quevedo, \emph{{Kahler moduli inflation}},
  \href{https://doi.org/10.1088/1126-6708/2006/01/146}{\emph{JHEP} {\bfseries
  01} (2006) 146} [\href{https://arxiv.org/abs/hep-th/0509012}{{\ttfamily
  hep-th/0509012}}].

\bibitem{Cicoli:2008gp}
M.~Cicoli, C.~P. Burgess and F.~Quevedo, \emph{{Fibre Inflation: Observable
  Gravity Waves from IIB String Compactifications}},
  \href{https://doi.org/10.1088/1475-7516/2009/03/013}{\emph{JCAP} {\bfseries
  03} (2009) 013} [\href{https://arxiv.org/abs/0808.0691}{{\ttfamily
  0808.0691}}].

\bibitem{Baumann:2014nda}
D.~Baumann and L.~McAllister, \emph{{Inflation and String Theory}}, Cambridge
  Monographs on Mathematical Physics. Cambridge University Press, 5, 2015,
  \href{https://doi.org/10.1017/CBO9781316105733}{10.1017/CBO9781316105733},
  [\href{https://arxiv.org/abs/1404.2601}{{\ttfamily 1404.2601}}].

\bibitem{Wetterich:1987fm}
C.~Wetterich, \emph{{Cosmology and the Fate of Dilatation Symmetry}},
  \href{https://doi.org/10.1016/0550-3213(88)90193-9}{\emph{Nucl. Phys. B}
  {\bfseries 302} (1988) 668}
  [\href{https://arxiv.org/abs/1711.03844}{{\ttfamily 1711.03844}}].

\bibitem{Copeland:1997et}
E.~J. Copeland, A.~R. Liddle and D.~Wands, \emph{{Exponential potentials and
  cosmological scaling solutions}},
  \href{https://doi.org/10.1103/PhysRevD.57.4686}{\emph{Phys. Rev. D}
  {\bfseries 57} (1998) 4686}
  [\href{https://arxiv.org/abs/gr-qc/9711068}{{\ttfamily gr-qc/9711068}}].

\bibitem{Ferreira:1997hj}
P.~G. Ferreira and M.~Joyce, \emph{{Cosmology with a primordial scaling
  field}}, \href{https://doi.org/10.1103/PhysRevD.58.023503}{\emph{Phys. Rev.
  D} {\bfseries 58} (1998) 023503}
  [\href{https://arxiv.org/abs/astro-ph/9711102}{{\ttfamily
  astro-ph/9711102}}].

\bibitem{Barreiro:1999zs}
T.~Barreiro, E.~J. Copeland and N.~J. Nunes, \emph{{Quintessence arising from
  exponential potentials}},
  \href{https://doi.org/10.1103/PhysRevD.61.127301}{\emph{Phys. Rev. D}
  {\bfseries 61} (2000) 127301}
  [\href{https://arxiv.org/abs/astro-ph/9910214}{{\ttfamily
  astro-ph/9910214}}].

\bibitem{Collinucci:2004iw}
A.~Collinucci, M.~Nielsen and T.~Van~Riet, \emph{{Scalar cosmology with
  multi-exponential potentials}},
  \href{https://doi.org/10.1088/0264-9381/22/7/005}{\emph{Class. Quant. Grav.}
  {\bfseries 22} (2005) 1269}
  [\href{https://arxiv.org/abs/hep-th/0407047}{{\ttfamily hep-th/0407047}}].

\bibitem{Shiu:2023nph}
G.~Shiu, F.~Tonioni and H.~V. Tran, \emph{{Accelerating universe at the end of
  time}}, \href{https://doi.org/10.1103/PhysRevD.108.063527}{\emph{Phys. Rev.
  D} {\bfseries 108} (2023) 063527}
  [\href{https://arxiv.org/abs/2303.03418}{{\ttfamily 2303.03418}}].

\bibitem{Shiu:2023fhb}
G.~Shiu, F.~Tonioni and H.~V. Tran, \emph{{Late-time attractors and cosmic
  acceleration}},
  \href{https://doi.org/10.1103/PhysRevD.108.063528}{\emph{Phys. Rev. D}
  {\bfseries 108} (2023) 063528}
  [\href{https://arxiv.org/abs/2306.07327}{{\ttfamily 2306.07327}}].

\bibitem{Sonner:2006yn}
J.~Sonner and P.~K. Townsend, \emph{{Recurrent acceleration in dilaton-axion
  cosmology}}, \href{https://doi.org/10.1103/PhysRevD.74.103508}{\emph{Phys.
  Rev. D} {\bfseries 74} (2006) 103508}
  [\href{https://arxiv.org/abs/hep-th/0608068}{{\ttfamily hep-th/0608068}}].

\bibitem{Russo:2018akp}
J.~G. Russo and P.~K. Townsend, \emph{{Late-time Cosmic Acceleration from
  Compactification}},
  \href{https://doi.org/10.1088/1361-6382/ab0804}{\emph{Class. Quant. Grav.}
  {\bfseries 36} (2019) 095008}
  [\href{https://arxiv.org/abs/1811.03660}{{\ttfamily 1811.03660}}].

\bibitem{Russo:2022pgo}
J.~G. Russo and P.~K. Townsend, \emph{{A dilaton-axion model for string
  cosmology}}, \href{https://doi.org/10.1007/JHEP06(2022)001}{\emph{JHEP}
  {\bfseries 06} (2022) 001}
  [\href{https://arxiv.org/abs/2203.09398}{{\ttfamily 2203.09398}}].

\bibitem{Conlon:2022pnx}
J.~P. Conlon and F.~Revello, \emph{{Catch-me-if-you-can: the overshoot problem
  and the weak/inflation hierarchy}},
  \href{https://doi.org/10.1007/JHEP11(2022)155}{\emph{JHEP} {\bfseries 11}
  (2022) 155} [\href{https://arxiv.org/abs/2207.00567}{{\ttfamily
  2207.00567}}].

\bibitem{Apers:2022cyl}
F.~Apers, J.~P. Conlon, M.~Mosny and F.~Revello, \emph{{Kination, meet Kasner:
  on the asymptotic cosmology of string compactifications}},
  \href{https://doi.org/10.1007/JHEP08(2023)156}{\emph{JHEP} {\bfseries 08}
  (2023) 156} [\href{https://arxiv.org/abs/2212.10293}{{\ttfamily
  2212.10293}}].

\bibitem{Montero:2022prj}
M.~Montero, C.~Vafa and I.~Valenzuela, \emph{{The dark dimension and the
  Swampland}}, \href{https://doi.org/10.1007/JHEP02(2023)022}{\emph{JHEP}
  {\bfseries 02} (2023) 022}
  [\href{https://arxiv.org/abs/2205.12293}{{\ttfamily 2205.12293}}].

\bibitem{Giddings:2001yu}
S.~B. Giddings, S.~Kachru and J.~Polchinski, \emph{{Hierarchies from fluxes in
  string compactifications}},
  \href{https://doi.org/10.1103/PhysRevD.66.106006}{\emph{Phys. Rev. D}
  {\bfseries 66} (2002) 106006}
  [\href{https://arxiv.org/abs/hep-th/0105097}{{\ttfamily hep-th/0105097}}].

\bibitem{Cicoli:2011it}
M.~Cicoli, M.~Kreuzer and C.~Mayrhofer, \emph{{Toric K3-Fibred Calabi-Yau
  Manifolds with del Pezzo Divisors for String Compactifications}},
  \href{https://doi.org/10.1007/JHEP02(2012)002}{\emph{JHEP} {\bfseries 02}
  (2012) 002} [\href{https://arxiv.org/abs/1107.0383}{{\ttfamily 1107.0383}}].

\bibitem{Grimm:2019ixq}
T.~W. Grimm, C.~Li and I.~Valenzuela, \emph{{Asymptotic Flux Compactifications
  and the Swampland}},
  \href{https://doi.org/10.1007/JHEP06(2020)009}{\emph{JHEP} {\bfseries 06}
  (2020) 009} [\href{https://arxiv.org/abs/1910.09549}{{\ttfamily
  1910.09549}}].

\bibitem{Bastian:2021eom}
B.~Bastian, T.~W. Grimm and D.~van~de Heisteeg, \emph{{Modeling General
  Asymptotic Calabi-Yau Periods}},
  \href{https://arxiv.org/abs/2105.02232}{{\ttfamily 2105.02232}}.

\bibitem{Bastian:2021hpc}
B.~Bastian, T.~W. Grimm and D.~van~de Heisteeg, \emph{{Engineering small flux
  superpotentials and mass hierarchies}},
  \href{https://doi.org/10.1007/JHEP02(2023)149}{\emph{JHEP} {\bfseries 02}
  (2023) 149} [\href{https://arxiv.org/abs/2108.11962}{{\ttfamily
  2108.11962}}].

\bibitem{Grimm:2021ckh}
T.~W. Grimm, E.~Plauschinn and D.~van~de Heisteeg, \emph{{Moduli stabilization
  in asymptotic flux compactifications}},
  \href{https://doi.org/10.1007/JHEP03(2022)117}{\emph{JHEP} {\bfseries 03}
  (2022) 117} [\href{https://arxiv.org/abs/2110.05511}{{\ttfamily
  2110.05511}}].

\bibitem{Grimm:2022xmj}
T.~W. Grimm, S.~Lanza and T.~van Vuren, \emph{{Global symmetry-breaking and
  generalized theta-terms in Type IIB EFTs}},
  \href{https://doi.org/10.1007/JHEP10(2023)154}{\emph{JHEP} {\bfseries 10}
  (2023) 154} [\href{https://arxiv.org/abs/2211.11769}{{\ttfamily
  2211.11769}}].

\bibitem{Grana:2022dfw}
M.~Gra\~na, T.~W. Grimm, D.~van~de Heisteeg, A.~Herraez and E.~Plauschinn,
  \emph{{The tadpole conjecture in asymptotic limits}},
  \href{https://doi.org/10.1007/JHEP08(2022)237}{\emph{JHEP} {\bfseries 08}
  (2022) 237} [\href{https://arxiv.org/abs/2204.05331}{{\ttfamily
  2204.05331}}].

\bibitem{Bastian:2023shf}
B.~Bastian, D.~van~de Heisteeg and L.~Schlechter, \emph{{Beyond Large Complex
  Structure: Quantized Periods and Boundary Data for One-Modulus
  Singularities}},  \href{https://arxiv.org/abs/2306.01059}{{\ttfamily
  2306.01059}}.

\bibitem{vandeHeisteeg:2022gsp}
D.~T.~E. van~de Heisteeg, \emph{{Asymptotic String Compactifications: Periods,
  flux potentials, and the swampland}}, Ph.D. thesis, Utrecht U., 2022.
\newblock \href{https://arxiv.org/abs/2207.00303}{{\ttfamily 2207.00303}}.
\newblock 10.33540/1380.

\bibitem{Grimm:2020ouv}
T.~W. Grimm and C.~Li, \emph{{Universal axion backreaction in flux
  compactifications}},
  \href{https://doi.org/10.1007/JHEP06(2021)067}{\emph{JHEP} {\bfseries 06}
  (2021) 067} [\href{https://arxiv.org/abs/2012.08272}{{\ttfamily
  2012.08272}}].

\bibitem{Wiggins:2003}
Wiggins, \emph{{Introduction to Applied Nonlinear Dynamical Systems and
  Chaos}}. Springer New York, NY, 10, 2003,
  \href{https://doi.org/10.1017/9781108937092}{10.1017/9781108937092}.

\bibitem{Conlon:2021cjk}
J.~P. Conlon, S.~Ning and F.~Revello, \emph{{Exploring the holographic
  Swampland}}, \href{https://doi.org/10.1007/JHEP04(2022)117}{\emph{JHEP}
  {\bfseries 04} (2022) 117}
  [\href{https://arxiv.org/abs/2110.06245}{{\ttfamily 2110.06245}}].

\bibitem{Apers:2024ffe}
F.~Apers, J.~P. Conlon, E.~J. Copeland, M.~Mosny and F.~Revello, \emph{{String
  Theory and the First Half of the Universe}},
  \href{https://arxiv.org/abs/2401.04064}{{\ttfamily 2401.04064}}.

\bibitem{German:2001tz}
G.~German, G.~G. Ross and S.~Sarkar, \emph{{Low scale inflation}},
  \href{https://doi.org/10.1016/S0550-3213(01)00258-9}{\emph{Nucl. Phys. B}
  {\bfseries 608} (2001) 423}
  [\href{https://arxiv.org/abs/hep-ph/0103243}{{\ttfamily hep-ph/0103243}}].

\bibitem{Conlon:2008cj}
J.~P. Conlon, R.~Kallosh, A.~D. Linde and F.~Quevedo, \emph{{Volume Modulus
  Inflation and the Gravitino Mass Problem}},
  \href{https://doi.org/10.1088/1475-7516/2008/09/011}{\emph{JCAP} {\bfseries
  09} (2008) 011} [\href{https://arxiv.org/abs/0806.0809}{{\ttfamily
  0806.0809}}].

\bibitem{Bond:2006nc}
J.~R. Bond, L.~Kofman, S.~Prokushkin and P.~M. Vaudrevange, \emph{{Roulette
  inflation with Kahler moduli and their axions}},
  \href{https://doi.org/10.1103/PhysRevD.75.123511}{\emph{Phys. Rev. D}
  {\bfseries 75} (2007) 123511}
  [\href{https://arxiv.org/abs/hep-th/0612197}{{\ttfamily hep-th/0612197}}].

\bibitem{Hebecker:2014kva}
A.~Hebecker, P.~Mangat, F.~Rompineve and L.~T. Witkowski, \emph{{Tuning and
  Backreaction in F-term Axion Monodromy Inflation}},
  \href{https://doi.org/10.1016/j.nuclphysb.2015.03.015}{\emph{Nucl. Phys. B}
  {\bfseries 894} (2015) 456}
  [\href{https://arxiv.org/abs/1411.2032}{{\ttfamily 1411.2032}}].

\bibitem{Brustein:1992nk}
R.~Brustein and P.~J. Steinhardt, \emph{{Challenges for superstring
  cosmology}}, \href{https://doi.org/10.1016/0370-2693(93)90384-T}{\emph{Phys.
  Lett. B} {\bfseries 302} (1993) 196}
  [\href{https://arxiv.org/abs/hep-th/9212049}{{\ttfamily hep-th/9212049}}].

\bibitem{Barreiro:1998aj}
T.~Barreiro, B.~de~Carlos and E.~J. Copeland, \emph{{Stabilizing the dilaton in
  superstring cosmology}},
  \href{https://doi.org/10.1103/PhysRevD.58.083513}{\emph{Phys. Rev. D}
  {\bfseries 58} (1998) 083513}
  [\href{https://arxiv.org/abs/hep-th/9805005}{{\ttfamily hep-th/9805005}}].

\bibitem{Gouttenoire:2019kij}
Y.~Gouttenoire, G.~Servant and P.~Simakachorn, \emph{{Beyond the Standard
  Models with Cosmic Strings}},
  \href{https://doi.org/10.1088/1475-7516/2020/07/032}{\emph{JCAP} {\bfseries
  07} (2020) 032} [\href{https://arxiv.org/abs/1912.02569}{{\ttfamily
  1912.02569}}].

\bibitem{Co:2021lkc}
R.~T. Co, D.~Dunsky, N.~Fernandez, A.~Ghalsasi, L.~J. Hall, K.~Harigaya et~al.,
  \emph{{Gravitational wave and CMB probes of axion kination}},
  \href{https://doi.org/10.1007/JHEP09(2022)116}{\emph{JHEP} {\bfseries 09}
  (2022) 116} [\href{https://arxiv.org/abs/2108.09299}{{\ttfamily
  2108.09299}}].

\bibitem{Gouttenoire:2021wzu}
Y.~Gouttenoire, G.~Servant and P.~Simakachorn, \emph{{Revealing the Primordial
  Irreducible Inflationary Gravitational-Wave Background with a Spinning
  Peccei-Quinn Axion}},  \href{https://arxiv.org/abs/2108.10328}{{\ttfamily
  2108.10328}}.

\bibitem{Gouttenoire:2021jhk}
Y.~Gouttenoire, G.~Servant and P.~Simakachorn, \emph{{Kination cosmology from
  scalar fields and gravitational-wave signatures}},
  \href{https://arxiv.org/abs/2111.01150}{{\ttfamily 2111.01150}}.

\bibitem{Muia:2023wru}
F.~Muia, F.~Quevedo, A.~Schachner and G.~Villa, \emph{{Testing BSM physics with
  gravitational waves}},
  \href{https://doi.org/10.1088/1475-7516/2023/09/006}{\emph{JCAP} {\bfseries
  09} (2023) 006} [\href{https://arxiv.org/abs/2303.01548}{{\ttfamily
  2303.01548}}].

\bibitem{Sarangi:2002yt}
S.~Sarangi and S.~H.~H. Tye, \emph{{Cosmic string production towards the end of
  brane inflation}},
  \href{https://doi.org/10.1016/S0370-2693(02)01824-5}{\emph{Phys. Lett. B}
  {\bfseries 536} (2002) 185}
  [\href{https://arxiv.org/abs/hep-th/0204074}{{\ttfamily hep-th/0204074}}].

\bibitem{Copeland:2003bj}
E.~J. Copeland, R.~C. Myers and J.~Polchinski, \emph{{Cosmic F and D strings}},
  \href{https://doi.org/10.1088/1126-6708/2004/06/013}{\emph{JHEP} {\bfseries
  06} (2004) 013} [\href{https://arxiv.org/abs/hep-th/0312067}{{\ttfamily
  hep-th/0312067}}].

\bibitem{Hebecker:2023qke}
A.~Hebecker, S.~Schreyer and G.~Venken, \emph{{No asymptotic acceleration
  without higher-dimensional de Sitter vacua}},
  \href{https://doi.org/10.1007/JHEP11(2023)173}{\emph{JHEP} {\bfseries 11}
  (2023) 173} [\href{https://arxiv.org/abs/2306.17213}{{\ttfamily
  2306.17213}}].

\bibitem{VanRiet:2023cca}
T.~Van~Riet, \emph{{No accelerating scaling cosmologies at string tree
  level?}},  \href{https://arxiv.org/abs/2308.15035}{{\ttfamily 2308.15035}}.

\bibitem{Co:2019wyp}
R.~T. Co and K.~Harigaya, \emph{{Axiogenesis}},
  \href{https://doi.org/10.1103/PhysRevLett.124.111602}{\emph{Phys. Rev. Lett.}
  {\bfseries 124} (2020) 111602}
  [\href{https://arxiv.org/abs/1910.02080}{{\ttfamily 1910.02080}}].

\bibitem{Co:2019jts}
R.~T. Co, L.~J. Hall and K.~Harigaya, \emph{{Axion Kinetic Misalignment
  Mechanism}},
  \href{https://doi.org/10.1103/PhysRevLett.124.251802}{\emph{Phys. Rev. Lett.}
  {\bfseries 124} (2020) 251802}
  [\href{https://arxiv.org/abs/1910.14152}{{\ttfamily 1910.14152}}].

\bibitem{Chang:2019tvx}
C.-F. Chang and Y.~Cui, \emph{{New Perspectives on Axion Misalignment
  Mechanism}}, \href{https://doi.org/10.1103/PhysRevD.102.015003}{\emph{Phys.
  Rev. D} {\bfseries 102} (2020) 015003}
  [\href{https://arxiv.org/abs/1911.11885}{{\ttfamily 1911.11885}}].

\bibitem{GrillidiCortona:2015jxo}
G.~Grilli~di Cortona, E.~Hardy, J.~Pardo~Vega and G.~Villadoro, \emph{{The QCD
  axion, precisely}},
  \href{https://doi.org/10.1007/JHEP01(2016)034}{\emph{JHEP} {\bfseries 01}
  (2016) 034} [\href{https://arxiv.org/abs/1511.02867}{{\ttfamily
  1511.02867}}].

\end{thebibliography}\endgroup
\end{document}